\documentclass[%
reprint,
superscriptaddress,
 amsmath,amssymb,
 aps,
prx,
longbibliography,
raggedbottom,
]{revtex4-2}
\usepackage{formatting}
\begin{document}
\title{Theory of quasiparticle generation by microwave drives in superconducting qubits}
\author{Shoumik Chowdhury}
\email{shoumikc@mit.edu}
\affiliation{Research Laboratory of Electronics, Massachusetts Institute of Technology, Cambridge, Massachusetts 02139, USA}
\affiliation{Department of Electrical Engineering and Computer Science, Massachusetts Institute of Technology, Cambridge, Massachusetts 02139, USA}
\author{Max Hays}
\email{maxhays@mit.edu}
\affiliation{Research Laboratory of Electronics, Massachusetts Institute of Technology, Cambridge, Massachusetts 02139, USA}
\author{Shantanu R. Jha}
\affiliation{Research Laboratory of Electronics, Massachusetts Institute of Technology, Cambridge, Massachusetts 02139, USA}
\affiliation{Department of Electrical Engineering and Computer Science, Massachusetts Institute of Technology, Cambridge, Massachusetts 02139, USA}
\author{Kyle Serniak}
\affiliation{Research Laboratory of Electronics, Massachusetts Institute of Technology, Cambridge, Massachusetts 02139, USA}
\affiliation{MIT Lincoln Laboratory, Lexington, Massachusetts 02421, USA}
\author{\\ Terry P. Orlando}
\affiliation{Research Laboratory of Electronics, Massachusetts Institute of Technology, Cambridge, Massachusetts 02139, USA}
\affiliation{Department of Electrical Engineering and Computer Science, Massachusetts Institute of Technology, Cambridge, Massachusetts 02139, USA}
\author{Jeffrey A. Grover}
\affiliation{Research Laboratory of Electronics, Massachusetts Institute of Technology, Cambridge, Massachusetts 02139, USA}
\author{William D. Oliver}
\email{william.oliver@mit.edu}
\affiliation{Research Laboratory of Electronics, Massachusetts Institute of Technology, Cambridge, Massachusetts 02139, USA}
\affiliation{Department of Electrical Engineering and Computer Science, Massachusetts Institute of Technology, Cambridge, Massachusetts 02139, USA}
\affiliation{Department of Physics, Massachusetts Institute of Technology, Cambridge, Massachusetts 02139, USA}

% =====================================================
%%%%%%%%%%%%%%%%%%%%%%%%%%%%%%%%%%%%%%%
%%%%%%%%%%%%%% ABSTRACT %%%%%%%%%%%%%%
%%%%%%%%%%%%%%%%%%%%%%%%%%%%%%%%%%%%%%%
\date{\today}
\begin{abstract}
Microwave drives play a central role in the control of superconducting quantum circuits, enabling qubit gates, readout, and parametric interactions. As the drive frequencies are typically an order of magnitude smaller than (twice) the superconducting gap, it is generally assumed that such drives do not disturb the BCS ground state. However, sufficiently strong drives can activate multiphoton pair-breaking processes that generate quasiparticles (QPs) and result in qubit errors. In this work, we present a theoretical framework for calculating the rates of multiphoton-assisted pair-breaking transitions induced by charge or flux-coupled microwave drives. Through illustrative examples, we show that photon-assisted QP generation may affect novel high-frequency dispersive readout architectures, as well as Floquet-engineered superconducting circuits operating under strong driving
\end{abstract}
\maketitle

% =====================================================
%%%%%%%%%%%%%%%%%%%%%%%%%%%%%%%%%%%%%%%
%%%%%%%%%%%%%% MAIN TEXT %%%%%%%%%%%%%%
%%%%%%%%%%%%%%%%%%%%%%%%%%%%%%%%%%%%%%%

\section{Introduction\label{sec:1_Introduction}}
Over the last several decades, superconducting circuits have emerged as a promising hardware technology with applications in quantum computing, classical computing, and sensing~\cite{makhlin2001quantum, devoret2004superconducting, clarke2008superconducting, devoret2013superconducting, girvin2014circuit, sensing2017review}. These circuits operate at frequencies in the microwave domain and are therefore typically probed and controlled using microwave drives. In the context of quantum computing with superconducting qubits~\cite{krantz2019quantum, kjaergaard2020superconducting, blais2020quantum}, microwave drives are used to perform qubit gates and readout, and to activate parametric processes within a single mode or between distinct modes~\cite{blais2021circuit}. There is also a growing interest in utilizing strong always-on drives for Floquet engineering~\cite{goldman2014periodically,bukov2015universal, sameti2019floquet, oka2019floquet, gandon2022floquet}, i.e., fundamentally altering the circuit behavior via a drive, so that the effective emergent Hamiltonian has useful properties~\cite{zhang2019engineering, mundada2020floquet, nguyen2024programmable}, such as the protection of quantum information~\cite{kapitzonium2024, geier2024self, thibodeau2024floquet,lewellen2025frozonium, sellem2025dissipative, puri2017engineering}. 

For most quantum applications of superconducting circuits, the devices are operated at temperatures far below the critical temperature of the superconductor, and thus the system is expected to reside in the Bardeen-Cooper-Schrieffer (BCS) ground state, with quasiparticle (QP) excitations being exponentially suppressed. Experimentally, however, a significant population of QPs is consistently observed in superconducting devices, far exceeding the density expected from thermal equilibrium at typical cryostat temperatures ($\approx$ 20 mK) \cite{aumentado2004nonequilibrium, segall2004dynamics,martinis2009energy, wang2014measurement,aumentado2023quasiparticle,serniak2018hot, connolly2024coexistence}. When these nonequilibrium QPs tunnel across  Josephson junctions (JJs) in the circuit, they cause qubit errors such as relaxation, heating, and dephasing~\cite{catelani2011relaxation,catelani2012decoherence, lutchyn2005quasiparticle,glazman2021bogoliubov}, along with an associated change in the charge parity.

Although the origin of these nonequilibrium QPs was unclear for several decades, a few sources have been identified experimentally in recent years. These include ionizing radiation such as cosmic ray muons or environmental gamma rays~\cite{patel2017phonon, wilen2021correlated,vepsalainen2020impact,mcewen2022resolving,harrington2024synchronous}, phonon bursts from mechanical noise~\cite{anthony2024stress,kono2024mechanically}, and infrared (IR) photons originating from higher-temperature stages of the cryostat~\cite{barends2011minimizing, benevides2024quasiparticle, diamond2022distinguishing}. In particular, IR and millimeter-wave photons with characteristic energies above $2\Delta$, where $\Delta$ ($\approx h\times45$~GHz for aluminum) is the superconducting gap energy, have been shown to couple across JJs in qubit circuits and break Cooper pairs, inducing decoherence via the resulting pair-breaking transitions~\cite{houzet2019photon, diamond2022distinguishing, liu2024quasiparticle, leppakangas2013effects,de2013charge}. By contrast, since microwave photons applied for control and readout have energies much lower than $2\Delta$, it is generally assumed that they cannot break Cooper pairs or cause photon-assisted tunneling events. However, for sufficiently strong microwave drive powers, multiphoton processes could in principle supply the requisite energy $2\Delta$ needed to activate such pair-breaking transitions---a possibility we examine here.

In this work, we present a theoretical and numerical treatment of multiphoton-induced QP generation in JJ-based superconducting qubits. We develop an exact description of the driven qubit circuit using Floquet theory in Sec.~\ref{sec:2_Theoretical_Model}, and perturbatively treat the coupling to the QP degrees of freedom to calculate decoherence rates arising from pair-breaking transitions. Through worked examples in Secs.~\ref{sec:3_Charge_Coupled} and \ref{sec:4_Flux_Coupled}, we explore the conditions under which charge- and flux-coupled microwave drives induce pair-breaking. Our results establish fundamental limits on the use of strong driving in superconducting circuits, and provide guidelines for the design of next-generation readout techniques and Floquet-engineered qubits.

\section{Theoretical Model\label{sec:2_Theoretical_Model}}
\begin{figure}[t]
    \centering    \includegraphics[width=\linewidth]{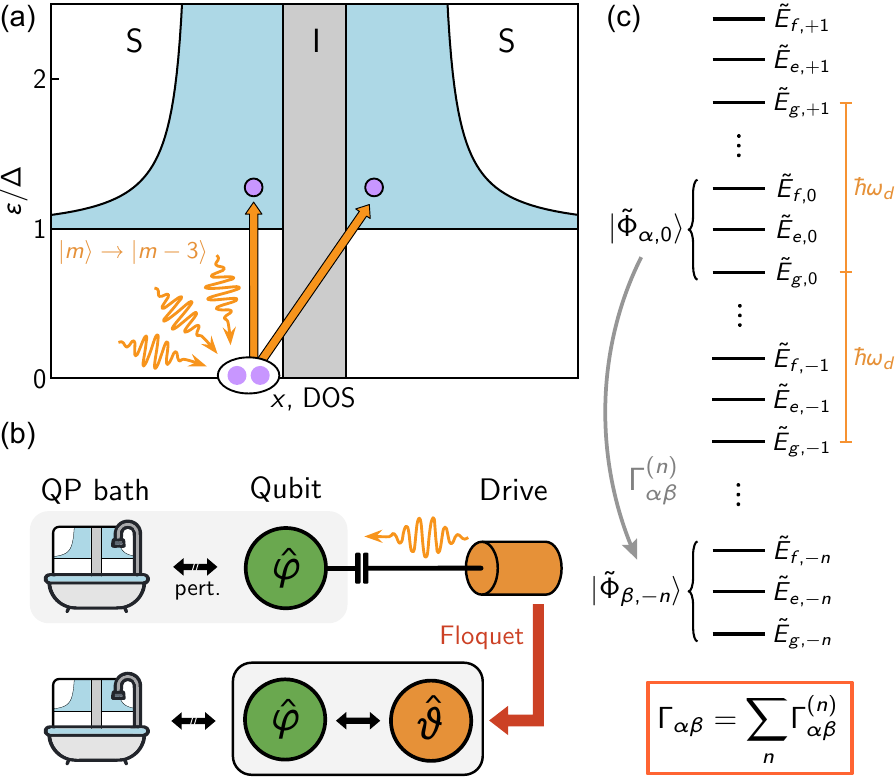}
    \caption{\textbf{Problem setup.} (a) Multiphoton-induced pair-breaking across a JJ in a driven superconducting qubit, where the energy $n\hbar\omega_d$ from $n$ drive photons is absorbed to break a Cooper pair and excite a pair of QPs above the gap. Here, an example with $n = 3$ is shown. (b) The system is modeled as a driven qubit coupled to a bath of BCS QPs. Using Floquet theory, we promote the drive to a fictitious quantum degree of freedom, such that resulting Floquet Hamiltonian is time-independent. (c) The spectrum of the Floquet Hamiltonian consists of infinite replicas of the dressed qubit energies, separated by multiples of $\hbar\omega_d$. We compute transition rates in the Floquet extended Hilbert space between the states $\ket{\tilde{\Phi}_{\alpha, 0}}$ and $\ket{\tilde{\Phi}_{\beta, -n}}$ using a Floquet-basis version of Fermi's golden rule, representing transitions between qubit states $\alpha \to \beta$ accompanied by the absorption of $n$ drive photons. The actual transition rates $\Gamma_{\alpha\beta}$ in the original qubit Hilbert space are then obtained via a summation over the different photon-number contributions in Floquet space. }
        \label{fig:1_schematic}
\end{figure}

As depicted in Fig. \ref{fig:1_schematic}, we consider a driven superconducting qubit coupled to a bath of QPs \cite{catelani2011relaxation, glazman2021bogoliubov}. The system dynamics are governed by the Hamiltonian
\begin{equation}
    \hat{H}(t) = \hat{H}_q(t) + \hat{H}_{\rm QP} + \hat{H}_T(t).
\label{eq:1_general_qubit_QP_coupling}
\end{equation}
\noindent Here, $\hat{H}_q(t)$ and $\hat{H}_{\rm QP}$ are the Hamiltonians of the qubit and QPs, respectively, whereas $\hat{H}_T(t)$ captures the coupling between them and will be treated as a perturbation. The qubit and coupling terms above are time periodic with the period $\mathbb{T}$ of the microwave drive: $\hat{H}_q(t) = \hat{H}_q(t+\mathbb{T})$ and $\hat{H}_T(t) = \hat{H}_T(t+\mathbb{T})$. 
We aim to calculate the rate at which $H_T(t)$ induces pair-breaking transitions between the time-dependent eigenstates of $\hat{H}_q(t) + \hat{H}_{\rm QP}$. Although the theory and numerics we present here may, in principle, be applied to any JJ-based superconducting circuit driven by an arbitrary periodic signal, we explain our approach using the specific case of a single-junction transmon qubit~\cite{koch2007transmon} under a monochromatic charge drive with angular driving frequency $\omega_d = 2\pi/\mathbb{T}$. 
Anchored in this example, we begin by discussing how to construct the three terms of Eq.~\eqref{eq:1_general_qubit_QP_coupling}. 

First, consider $\hat{H}_q(t)$. In general, a microwave drive on a superconducting qubit will induce an oscillating voltage $V\cos(\omega_d t)$ across the JJ that couples to both the qubit and QP degrees of freedom; this would result in a time-dependent bath. Fortunately, the BCS calculation can be greatly simplified by setting up the full system Hamiltonian in a charge-displaced frame (see Appendix~\ref{app:A_BCS_Setup} for details). In this frame, $\hat{H}_{\rm QP}$ is rendered time-independent and the transmon qubit Hamiltonian $\hat{H}_q(t)$ takes the form
\begin{equation}
    \hat{H}_q(t) = 4E_C(\hat{n}-n_g)^2 - E_J\cos\left(\hat{\varphi} + \varphi_d\sin(\omega_d t)\right).
    \label{eq:H_transmon_disp_frame}
\end{equation}
Here, $E_J$ and $E_C$ are the Josephson and single-electron charging energies of the transmon, respectively, and $n_g$ is the dimensionless offset charge. The operator $\hat{\varphi}$ represents the phase drop across the JJ, and $\hat{n}$ is its conjugate charge. Finally, $\varphi_d = 2eV/\hbar\omega_d$ denotes the magnitude of the oscillating phase across the JJ due to the drive, which depends on the drive frequency $\omega_d$ and induced voltage amplitude $V$ through the second Josephson relation. When the drive is turned off (i.e., $\varphi_d \to 0$), we recover the static transmon Hamiltonian, with eigenenergies and eigenstates labeled by $E_\alpha$ and $|\alpha\rangle$, respectively. 

The next term in Eq.~\eqref{eq:1_general_qubit_QP_coupling}, $\hat{H}_{\rm QP}$, is the BCS Hamiltonian for the QPs in the left and right junction electrodes, written in the excitation picture. This takes the form
\begin{equation}
    \hat{H}_{\rm QP} = \sum_{l, s} \varepsilon_l \hat{\gamma}_{ls}^\dagger \hat{\gamma}_{ls} + \sum_{r, s} \varepsilon_r \hat{\gamma}_{rs}^\dagger \hat{\gamma}_{rs}.
\label{eq:H_QP_bath}
\end{equation}
The labels $l$ and $r$ index QP states on the left and right sides of the JJ, respectively, and $s \in \{\uparrow, \downarrow\}$ indexes their spin. We thus read $\hat{\gamma}_{ls}$ ($\hat{\gamma}_{rs}$) as the fermionic annihilation operator for a QP in the left (right) lead of the JJ.
The QP energies $\varepsilon_k = \sqrt{\xi_k^2 + \Delta^2}$ are written in terms of the normal-state electron energy $\xi_k$ (measured from the chemical potential) and the superconducting gap $\Delta$.
For simplicity, we consider the gap to be identical on both sides of the JJ throughout this work, though we note that it would be straightforward to extend our treatment to the experimentally relevant case of asymmetric gaps~\cite{aumentado2004nonequilibrium,diamond2022distinguishing, marchegiani2022asymmetric, mcewen2024resisting,harrington2024synchronous, nho2025recovery}. In all presented calculations, we use the superconducting gap for (bulk) aluminum: $2\Delta/h = 90$ GHz. 

Finally, the last term, $\hat{H}_T$, in Eq.~\eqref{eq:1_general_qubit_QP_coupling} couples the qubit and QP degrees of freedom. It models electrons tunneling across the junction:
\begin{align}
    \begin{split}
        \hat{H}_T(t) = \tau \sum\limits_{l, r, s} &\big[e^{i(\hat{\varphi} + \varphi_d(t))/2} \hat{c}_{rs}^\dagger \hat{c}_{ls} + {\rm h.c.}\big] \\ &+ E_J\cos(\hat{\varphi} + \varphi_d(t)).
    \label{eq:H_T_electron_tunneling}
    \end{split}
\end{align}
Here the operators $\hat{c}_{ks} = u_k\hat{\gamma}_{ks} + \sigma(s) v_k\hat{\gamma}_{k\bar{s}}^\dagger$ are the annihilation operators for electrons, constructed via a Bogoliubov transformation, where $\sigma(s) = \pm 1$ for $s = \,\uparrow, \downarrow$ and $\bar{s}$ denotes the opposite spin to $s$.
The conventional BCS coherence factors $u_k, v_k$ are defined via $|v_k|^2 = 1 - |u_k|^2 = (1 - \xi_k/\varepsilon_k)/2$ \cite{glazman2021bogoliubov}. We can also express the small tunneling amplitude $\tau$ in terms of the qubit Josephson energy $E_J$ by using the Ambegaokar-Baratoff relation~\cite{ambegaokar1963tunneling}: $\tau^2 = (e^2/\pi\hbar)\cdot E_J / \Delta\pi^2 \nu_0^2$, where $\nu_0$ is the density of states per spin of normal electrons at the Fermi level. Note that, here, $\hat{H}_T(t)$ explicitly depends on the drive-induced phase $\varphi_d(t) = \varphi_d\sin(\omega_d t)$. In addition, the term  $E_J\cos(\hat{\varphi} + \varphi_d(t))$ is added here to avoid double counting the Josephson energy part of the qubit Hamiltonian, which we already included in Eq.~\eqref{eq:H_transmon_disp_frame}; this is formally important, since the Josephson term itself arises from performing second-order perturbation theory on the first line of $\hat{H}_T$ with small parameter $\tau$~\cite{supp}. We can drop this double-counting correction term going forward, however, as it will not contribute to any QP transitions when treating $\hat{H}_T(t)$ as a perturbation in Fermi's golden rule. 

If we rewrite the tunneling Hamiltonian using the QP operators above, we obtain an expression with two terms: 
\begin{align}
\begin{split}
    &\!\!\hat{H}_T(t) = \\ &\!\!\tau \sum_{l, r, s}\Big[(u_ru_l e^{i(\hat{\varphi} + \varphi_d(t))/2} - v_rv_l e^{-i(\hat{\varphi} + \varphi_d(t))/2})\hat{\gamma}_{r s}^{\dagger} \hat{\gamma}_{ls} \,\, + \\ \, &\!\!\sigma(s)(u_rv_l e^{i(\hat{\varphi} + \varphi_d(t))/2} + v_ru_l e^{-i(\hat{\varphi} + \varphi_d(t))/2})\hat{\gamma}_{r s}^{\dagger} \hat{\gamma}_{l \bar{s}}^{\dagger}\Big] + {\rm h.c.} 
\end{split}
\label{eq:H_T_QP_and_PAT}
\end{align}
The first term ($\hat{\gamma}_{r s}^{\dagger} \hat{\gamma}_{l s}$) conserves the total number of QPs and thus describes the conventional process of QP tunneling across the junction \cite{catelani2011relaxation,catelani2012decoherence,glazman2021bogoliubov}, whereas the second term ($\hat{\gamma}_{r s}^{\dagger} \hat{\gamma}_{l \bar{s}}^{\dagger}$) models photon-assisted QP generation \cite{houzet2019photon,kyle2019thesis}, a process which breaks Cooper pairs and places the two resulting QPs on either side of the JJ [see Fig.~\ref{fig:1_schematic}(a)]. This generation process requires the absorption of energy larger than twice the gap ($\geq 2\Delta$) and can occur even if there are no pre-existing QPs in the device. By contrast, the rate of any state transitions arising from the tunneling term will be proportional to $x_{\rm qp} = n_{\rm qp}/n_{\rm cp}$, defined as the density $n_{\rm qp}$ of QPs normalized by the density $n_{\rm cp}$ of Cooper pairs (with $x_{\rm qp} \sim 10^{-8}$ to $10^{-5}$ in typical devices \cite{aumentado2004nonequilibrium, segall2004dynamics,wang2014measurement,aumentado2023quasiparticle,serniak2018hot}). We briefly discuss how the drive can affect the conventional QP tunneling process in Appendix~\ref{app:C_Conventional_QPT}. However, since our primary focus here is on QP generation, we drop the $\hat{\gamma}_{r s}^{\dagger} \hat{\gamma}_{l s}$ term above going forward (i.e., effectively setting $x_{\rm qp} = 0$) and consider only the second term in Eq.~\eqref{eq:H_T_QP_and_PAT}, which we can rewrite as
\begin{align}
\begin{split}
    &\!\!\!\hat{H}_T(t) \,=\,\, \tau \sum_{l, r, s}\sigma(s)\bigg[\left(u_r v_l+v_r u_l\right) \cos\bigg(\frac{\hat{\varphi} + \varphi_d(t)}{2}\bigg) \\ &\quad\,\,\,\,+i\left(u_r v_l-v_r u_l\right) \sin \bigg(\frac{\hat{\varphi} + \varphi_d(t)}{2}\bigg)\bigg]\hat{\gamma}_{r s}^{\dagger} \hat{\gamma}_{l \bar{s}}^{\dagger} + {\rm h.c.}
    \label{eq:H_PAT_only}
\end{split}
\end{align}
Having constructed the three terms of Eq.~\eqref{eq:1_general_qubit_QP_coupling}, we now proceed toward our goal of calculating the rate at which $\hat{H}_T(t)$ induces transitions between the time-dependent eigenstates of the driven system. In this work, we consider strongly driven superconducting qubits, where the pair-breaking energy is supplied by microwave photons from the drive via multiphoton absorption. In this limit, the effect of the drive on the qubit must be treated nonperturbatively. 

To this end, we use the extended Hilbert space formulation of Floquet theory, also referred to as the replicas picture, to exactly model the driven system~\cite{shirley1965solution, sambe1973steady, grifoni1998driven, drese1999floquet, eckardt2015high, di2022extensible, rudner2020floquet, kitagawa2011transport, bilitewski2015scattering, faisal1987theory}. Specifically, we can expand the time-dependent qubit Hamiltonian $\hat{H}_q(t)$ into a time-independent Floquet Hamiltonian $\hat{H}_q^F$ which exists in a larger, fictitious Hilbert space  (see Appendix~\ref{app:B_Floquet_Theory}). To make the conversion, we first rewrite $\hat{H}_q(t)$ in terms of the phase of the drive $\theta(t) = \omega_d t$, and then promote this quantity to a $2\pi$-periodic quantum degree of freedom $\hat{\vartheta}$ with conjugate variable $\hat{m}$. Effectively, we are dealing with a coupled two-mode Hamiltonian involving the bare qubit mode and the drive mode. The extended-space Floquet Hamiltonian for a transmon under a charge drive [cf. Eq.~\eqref{eq:H_transmon_disp_frame}] is then
\begin{align}
    \begin{split}
        \!\!\!\hat{H}_q^F &\equiv \hat{H}_q(t) - i\hbar\!\partial_t \\ &= 4E_C(\hat{n}-n_g)^2 - E_J\cos(\hat{\varphi} + \varphi_d\sin\hat{\vartheta})  + \hbar\omega_d\hat{m}.\!\!
    \end{split}
\label{eq:floquet_effective_HF_qubit}
\end{align}
Here, we express the operators ${\hat{m} = \sum_m m\op{m}{m}}$ and ${\sin\hat{\vartheta} = \frac{1}{2i}\sum_m \op{m+1}{m} + {\rm h.c.}}$ in the $\hat{m}$-basis, with $m \in \mathbb{Z}$.
When the drive is turned off (i.e., $\varphi_d \to 0$), the Floquet eigenstates $\ket{\Phi_{\alpha, m}}$ of $\hat{H}_q^F$ are given by product states $\ket{\Phi_{\alpha, m}}\equiv\ket{\alpha}\ket{m}$ with energies $E_\alpha + m\hbar\omega_d$, corresponding to an infinite set of replicas of the original qubit spectrum separated by multiples of $\hbar\omega_d$.
As the drive is turned on, the coupling term causes the qubit and drive degrees of freedom to hybridize, resulting in Floquet eigenstates $\ket{\tilde{\Phi}_{\alpha, m}} \equiv \ket{\widetilde{\alpha, m}}$ that are dressed states of the two modes.
The associated dressed energies $\tilde{E}_{\alpha, m}$ vary with drive strength via the ac-Stark shift, and thus we can have multiphoton resonances $\tilde{E}_{\alpha, m} \approx \tilde{E}_{\beta, n}$ between the states $\ket{\tilde{\Phi}_{\alpha, m}}$ and $\ket{\tilde{\Phi}_{\beta, n}}$. In the original qubit Hilbert space, these correspond to transitions between qubit states $\alpha$ and $\beta$ accompanied by the absorption or emission of $|n-m|$ drive quanta.
For this reason, we interpret the drive quantum number as the number of photons exchanged between the qubit and drive.
One challenge faced when performing Floquet simulations is the ambiguity in labeling the dressed eigenstates. In our simulations, we use a highly efficient labeling algorithm inspired from Refs. \cite{shillito2022dynamics,xiao2023diagrammatic,dumas2024measurement}, developed using the \texttt{jaxquantum} numerical simulation library \cite{jha2024jaxquantum}. We discuss this procedure and provide a concrete example in Appendix~\ref{app:B_Floquet_Theory}. 

Using the Floquet extended Hilbert space, we can now rewrite the full driven-qubit--QP Hamiltonian in the Floquet basis by substituting $\hat{H}_q(t)\to\hat{H}_q^F$ in Eqs. (\ref{eq:1_general_qubit_QP_coupling}--\ref{eq:H_transmon_disp_frame}) and similarly promoting the perturbation Hamiltonian to the extended Hilbert space, $\hat{H}_T(t) \to \hat{H}_T^F$. Since the full system Hamiltonian is now time-independent in the extended space, we can set up a Floquet Fermi's golden rule calculation using standard tools from time-independent perturbation theory \cite{faisal1987theory, kitagawa2011transport, bilitewski2015scattering, rudner2020floquet}. Specifically, we compute the transition rate between qubit states $\alpha$ and $\beta$ with a corresponding absorption of energy $n\hbar\omega_d$ (i.e, due to $n$ drive photons) in the superconductor, leading to the creation of a pair of QP excitations:  
\begin{align}
    \begin{split}
         \Gamma_{\alpha\beta}^{(n)} &=\frac{2\pi}{\hbar} \sum_{l, r, s}\big|\big\langle{\rm GS}\big|\big\langle \tilde{\Phi}_{\beta, -n}\big|\hat{\gamma}_{r s}\hat{\gamma}_{l \bar{s}}\hat{H}_T^F\big|\tilde{\Phi}_{\alpha, 0}\big\rangle\big|{\rm GS}\big\rangle\big|^2 \\ &\qquad\qquad\times\delta(\hbar\omega_{\alpha\beta}^{(n)} -\varepsilon_l-\varepsilon_r).
    \end{split}
    \label{eq:Gamma_abn}
\end{align}
Here $\ket{\rm GS}$ denotes the BCS ground state of the two leads, and $\hbar\omega_{\alpha\beta}^{(n)} = \tilde{E}_{\alpha, 0} - \tilde{E}_{\beta, -n} = n\hbar\omega_d + \tilde{E}_{\alpha, 0} - \tilde{E}_{\beta, 0}$ is the energy difference between the initial and final extended-space Floquet states. We note that the choice of Floquet photon-number index $m=0$ for the initial state is arbitrary here: we could also equally well use $\ket{\tilde{\Phi}_{\alpha, m}}$ and $\ket{\tilde{\Phi}_{\beta, m-n}}$ (i.e., arbitrary $m$) as the initial and final states, since only differences in the photon-number index are relevant in the replicas picture. Upon evaluating Eq.~\eqref{eq:Gamma_abn}, we find that the resulting expression conveniently factorizes into
\begin{align}
         \!\!\!\Gamma_{\alpha\beta}^{(n)} &= \nonumber \\ &\!\!\!\!\!\!\!\!\Gamma_{\rm ph}\bigg[\big|\bra{\tilde{\Phi}_{\beta, -n}}\cos\bigg(\frac{\hat{\varphi} + \varphi_d\sin\hat{\vartheta}}{2}\bigg)\ket{\tilde{\Phi}_{\alpha, 0}} \big|^2 \, \mathcal{S}_{\rm ph}^{+}\big(\omega_{\alpha\beta}^{(n)}\big) \nonumber \\
         &\!\!\!\!\!\!+\big|\bra{\tilde{\Phi}_{\beta, -n}}\sin\bigg(\frac{\hat{\varphi} + \varphi_d\sin\hat{\vartheta}}{2}\bigg)\ket{\tilde{\Phi}_{\alpha, 0}} \big|^2 \mathcal{S}_{\rm ph}^{-}\big(\omega_{\alpha\beta}^{(n)}\big)\bigg].
    \label{eq:Gamma_abn_eval}
\end{align}
The prefactor here is $\Gamma_{\rm ph} = 16E_J/h$, and the structure factors $\mathcal{S}_{\rm ph}^{\pm}$ for photon-assisted QP generation are defined as~\cite{houzet2019photon}
\begin{equation}
    \mathcal{S}_{\rm ph}^{\pm}(\omega) = \int_{1}^\infty\!\!\int_{1}^\infty dx dy \frac{xy \pm 1}{\sqrt{x^2-1}\sqrt{y^2-1}}\delta\bigg(\frac{\hbar\omega}{\Delta} - x - y\bigg),
    \label{eq:PAT_S_factors}
\end{equation}
with the integrals performed over dimensionless variables $x, y$ \footnote{The structure factors in Eq.~\eqref{eq:PAT_S_factors} can be expressed analytically in terms of complete elliptic integral functions:
\begin{equation*}
\quad\quad \mathcal{S}_{\rm ph}^{\pm}(\omega) = (z + 2) E\Big(\frac{z - 2}{z + 2}\Big) + 4\frac{z + 1 \pm 1}{z + 2} K\Big(\frac{z - 2}{z + 2}\Big),
\end{equation*}
where we have used the shorthand notation $z = \hbar\omega/\Delta$.}. Finally, in order to obtain the actual transition rates between qubit eigenstates $\alpha$ and $\beta$ in the original Hilbert space, we sum over the photon-number index \cite{bilitewski2015scattering}:
\begin{equation}
\Gamma_{\alpha\beta} = \sum_n \Gamma_{\alpha\beta}^{(n)}.
\label{eq:Gamma_ab_eval}
\end{equation}
This effectively amounts to a projection from the Floquet extended Hilbert space back onto the qubit Hilbert space. Equations~(\ref{eq:Gamma_abn_eval}--\ref{eq:Gamma_ab_eval}) represent the main theoretical results of this work: they express how the rates of pair-breaking transitions are calculated in the Floquet basis in terms of the dressed matrix elements of $\cos[(\hat{\varphi} + \varphi_d\sin\hat{\vartheta})/2]$ and $\sin[(\hat{\varphi} + \varphi_d\sin\hat{\vartheta})/2]$. From here, we also define the total parity switching rate for qubit state $\alpha$ by summing over all possible final qubit states $\beta$, i.e.,  $\Gamma_\alpha = \sum_\beta \Gamma_{\alpha\beta}$. The equivalent charge-parity lifetime is then $T_\alpha = 1/\Gamma_\alpha$.

\section{Charge-Coupled Drives\label{sec:3_Charge_Coupled}}
\subsection{Driven transmon qubits\label{subsec:drive_qubits}}
\begin{figure}[t]
    \centering    
    \includegraphics[width=\linewidth]{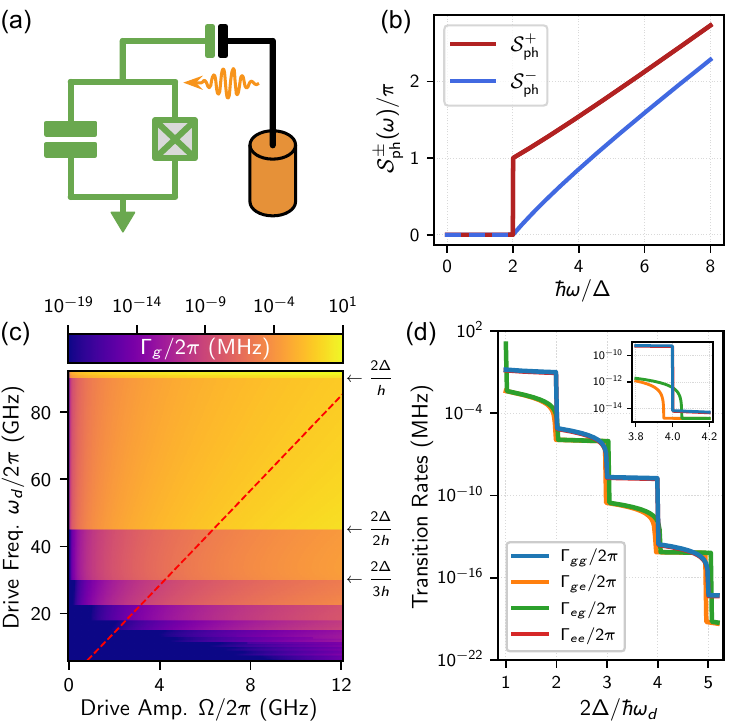}
    \caption{\textbf{Photon-assisted QP generation in transmon qubits.} (a) Circuit of a transmon qubit subjected to a microwave drive. (b) Structure factors $\mathcal{S}_{\rm ph}^\pm$ [Eq.~\eqref{eq:PAT_S_factors}] for photon-assisted QP generation  as a function of transition frequency. $\mathcal{S}_{\rm ph}^+$ and $\mathcal{S}_{\rm ph}^-$ have a threshold behavior at frequency $2\Delta/\hbar$. (c) Simulation of the total parity switching rate $\Gamma_g$ for the driven ground state of a transmon with $E_J/h = 3.025$ GHz, $E_C/h = 56$ MHz, and $n_g = 0$, as a function of the frequency $\omega_d$ and amplitude $\Omega$ of the drive. The rates exhibit sharp steps at $\hbar\omega_d = 2\Delta/n$, for integer $n$, reflecting the different $n$-photon processes required for pair-breaking. The red dashed line is a line of constant qubit ac-Stark shift $\delta_{\rm ac}$ (see the text). (d) Transition rates $\Gamma_{\alpha\beta}$ in the qubit $\{g, e\}$ manifold as a function of $2\Delta/\hbar\omega_d$ along the line of constant ac-Stark shift from panel (c). These rates display distinct steps, with alternating ``sharpness'' that depends on the even-odd parity of $n$, where $n = \lceil 2\Delta/\hbar\omega_d \rceil$ for state-preserving transitions ($\Gamma_{gg}, \Gamma_{ee}$) and $n = \lceil (2\Delta \pm\hbar\tilde{\omega}_q)/\hbar\omega_d \rceil$ for state-changing transitions ($\Gamma_{ge}$, $\Gamma_{eg}$) for Stark-shifted qubit frequency $\tilde{\omega}_q$. The inset shows an enlarged view of the behavior around the step $n = 4$.}
    \label{fig:2_transmon}
\end{figure}

We consider a charge-driven transmon depicted in Fig. \ref{fig:2_transmon}(a). The Hamiltonian for this qubit is typically written as \cite{koch2007transmon, krantz2019quantum, blais2021circuit}
\begin{equation}
    \hat{H}_q(t) = 4E_C(\hat{n}-n_g)^2 - E_J\cos(\hat{\varphi}) + \hbar\Omega\hat{n}\cos(\omega_d t),
    \label{eq:H_transmon_Rabi_frame}
\end{equation}
where the qubit parameters are the same as in Eq.~\eqref{eq:H_transmon_disp_frame}, and $\Omega$ denotes the drive amplitude. As mentioned above, however, we need to transform this Hamiltonian to a charge-displaced frame in order to be compatible with the framework presented in Sec.~\ref{sec:2_Theoretical_Model} (also see Appendix~\ref{app:A_BCS_Setup} for details on how Eq.~\eqref{eq:H_transmon_Rabi_frame} can be constructed from BCS theory). To this end, we apply a time-dependent unitary transformation $\hat{U}(t) = \exp(i\hbar\Omega\hat{n}\sin(\omega_d t)/\hbar\omega_d)$ to Eq.~\eqref{eq:H_transmon_Rabi_frame}, such that the resulting Hamiltonian $\hat{H}_q \to \hat{U}\hat{H}_q \hat{U}^\dagger + i\hbar[\partial_t\hat{U}]\hat{U}^\dagger$ is given by Eq.~\eqref{eq:H_transmon_disp_frame}, with an oscillating phase amplitude of $\varphi_d = \Omega/\omega_d$.

In Fig.~\ref{fig:2_transmon}(c), we numerically simulate the parity switching rate $\Gamma_g$ for the ground state of a transmon as a function of the amplitude $\Omega$ and frequency $\omega_d$ of an off-resonant charge drive. Using our state-labeling procedure (cf. Appendix~\ref{app:B_Floquet_Theory}), we identify the driven ground state $\ket{\tilde{\Phi}_{g, 0}}$ at each value of $\Omega$ and $\omega_d$ and compute transition rates to all other eigenstates, before finally summing over the final states to obtain $\Gamma_g$. The plot exhibits clear steps in the parity switching rate as a function of drive frequency, with the steps occurring here at fractions of $2\Delta$, i.e., whenever $\hbar\omega_d = 2\Delta/n$ for integer $n$. The steps reflect the threshold behavior of the structure factors $\mathcal{S}_{\rm ph}^\pm$ [Fig.~\ref{fig:2_transmon}(b)], and correspond to $n$-photon pair-breaking processes that become progressively weaker with increasing $n$. For typical charge driving parameters with $\Omega/2\pi \ll 1$ GHz and $\omega_d/2\pi < 10$ GHz, the pair-breaking rates are negligible [see lower left region of Fig.~\ref{fig:2_transmon}(c)] and so we do not expect this decoherence mechanism to be limiting (the rate $\Gamma_e$ for the qubit excited state, not shown, exhibits similar behavior). However, for sufficiently large drive amplitudes or when using higher frequency drives, decoherence from multiphoton pair-breaking becomes significant. 

To further visualize the steplike behavior, in Fig.~\ref{fig:2_transmon}(d) we separate the total parity switching rates $\Gamma_\alpha$ into their constituent transition rates $\Gamma_{\alpha\beta}$, focusing on transitions within the qubit $\{g, e\}$ manifold \footnote{In simulations with fixed-frequency transmons, we typically consider 20-30 qubit levels after moving into the energy eigenbasis; thus, the total parity switching rates $\Gamma_\alpha$ include contributions to all higher states. Given the structure of the transmon $\cos(\hat{\varphi}/2)$ and $\sin(\hat{\varphi}/2)$ operators used to calculate the transition rates, however, we find that the dominant rates $\Gamma_{\alpha\beta}$ occur for $\beta = \alpha\pm 1$.}. We plot $\Gamma_{gg}$, $\Gamma_{ge}$, $\Gamma_{eg}$, and $\Gamma_{ee}$ versus $2\Delta/\hbar\omega_d$, which approximately counts the number of photons absorbed in a pair-breaking transition. At each drive frequency $\omega_d$, we adjust the drive amplitude $\Omega$ to maintain a constant qubit ac-Stark shift $\delta_{\rm ac} = \tilde{\omega}_q - \omega_q$, where $\hbar\omega_q \equiv E_e - E_g$ and $\hbar\tilde{\omega}_q \equiv \tilde{E}_{e, 0} - \tilde{E}_{g, 0}$ denote the bare and Stark-shifted qubit energies, respectively; in Fig.~\ref{fig:2_transmon}(d), $\delta_{\rm ac}/2\pi = 3$ MHz. Since $\delta_{\rm ac}$ quantifies the effective strength of the drive, keeping it constant allows us to compare results at different drive frequencies in a consistent manner. For the state-preserving transitions ($\Gamma_{gg}, \Gamma_{ee}$) in Fig.~\ref{fig:2_transmon}(d), the steps occur at $n = \lceil 2\Delta/\hbar\omega_d \rceil$; for the state-changing transitions ($\Gamma_{ge}, \Gamma_{eg}$), the steps instead occur at $n = \lceil (2\Delta\pm\hbar\tilde{\omega}_q)/\hbar\omega_d \rceil$ (see also Ref. \cite{vlad2025qps}), reflecting energy given to or taken from the transmon. All transitions involve a change in the charge-parity and the creation of a pair of QPs in the device. A key feature of the steplike behavior of the transition rates is that the ``sharpness'' of the steps alternates depending on whether $n$ is even or odd. This alternating photon-number-parity effect can be interpreted using Eq.~\eqref{eq:Gamma_abn_eval} and originates from the distinct threshold behavior of the two structure factors: $\mathcal{S}_{\rm ph}^+(\omega)$ turns on abruptly at $\hbar\omega = 2\Delta$, producing sharp steps, whereas $\mathcal{S}_{\rm ph}^-(\omega)$ turns on gradually and continuously from zero, leading to rounded steps. Focusing on the first term of Eq.~\eqref{eq:Gamma_abn_eval} associated with $\mathcal{S}_{\rm ph}^+$ and writing $\textstyle \hat{\varphi}_d \equiv \varphi_d\sin\hat{\vartheta}$ for brevity, we can expand the cosine to get $\cos(\hat{\varphi}/2)\cos(\hat{\varphi}_d/2) - \sin(\hat{\varphi}/2)\sin(\hat{\varphi}_d/2)$. As $\cos(\hat{\varphi}/2)$ and $\sin(\hat{\varphi}/2)$ generate state-preserving and state-changing transitions, respectively, we find that the sharp steps will occur at even $n$ for state-preserving transitions and at odd $n$ for state-changing transitions (given that $\cos(\hat{\varphi}_d/2)$ and $\sin(\hat{\varphi}_d/2)$ contain only even and odd photon harmonics). Meanwhile, if we expand the sine in the second term of Eq.~\eqref{eq:Gamma_abn_eval} associated with $\mathcal{S}_{\rm ph}^-$, we find that the rounded steps occur at odd $n$ for state-preserving transitions and at even $n$ for state-changing transitions.

\subsection{Detuned readout\label{subsec:detuned_readout}}

We now consider the example of transmon qubit readout, which is modeled by the Hamiltonian \cite{blais2004cavity,blais2021circuit}
\begin{align}
    \begin{split}
        \hat{H}_{\rm qr}(t) &= 4E_C(\hat{n}-n_g)^2 - E_J\cos(\hat{\varphi}) + \hbar\omega_r\hat{a}^\dagger\hat{a} \\ &\quad+ i\hbar g\hat{n}(\hat{a}^\dagger - \hat{a}) + i\epsilon(t) (\hat{a}^\dagger - \hat{a}) \sin(\omega_d t).
    \end{split}
    \label{eq:H_qubit_readout}
\end{align}
Here, the transmon parameters are the same as in Eq.~\eqref{eq:H_transmon_disp_frame}, $\omega_r$ is the bare readout resonator frequency, $g$ is the qubit-resonator capacitive coupling strength, $\epsilon(t)$ is the readout drive amplitude, and $\hat{a}$ is the annihilation operator associated with the resonator mode. During readout, the resonator is typically driven with a resonant tone, $\omega_d = \omega_r$, resulting in a large coherent state in the resonator with $\bar{n}$ photons on average. It has been shown in several prior works that the resonator can be accurately modeled using a semi-classical picture \cite{shillito2022dynamics, cohen2023reminiscence,dumas2024measurement}. By moving Eq.~\eqref{eq:H_qubit_readout} to a displaced frame of the resonator and then tracing out the resonator degree of freedom, the resulting driven-qubit Hamiltonian is given by Eq.~\eqref{eq:H_transmon_Rabi_frame} with an equivalent drive amplitude of $\Omega = 2g\sqrt{\bar{n}}$. Finally, if we perform the same charge-displacement transformation from Sec.~\ref{subsec:drive_qubits}, we find that the readout problem can likewise be modeled using Eq.~\eqref{eq:H_transmon_disp_frame}, with the induced oscillating phase across the transmon JJ having an amplitude of $\varphi_d = 2g\sqrt{\bar{n}}/\omega_d$. Using this series of mappings enables us to apply the theoretical framework from Sec. \ref{sec:2_Theoretical_Model} to calculate the rates of multiphoton pair-breaking transitions in the presence of a readout drive. We note that for parameters that are typical for the  dispersive readout of superconducting qubits, the multiphoton-induced QP generation rate is small compared to other error channels. However, as we describe below, novel high-frequency readout schemes may need to consider these effects.

Recently, it was shown that detuning the readout resonator frequency from that of the transmon qubit can significantly improve readout fidelity~\cite{kurilovich2025high} and reduce the prevalence of measurement-induced state transitions (MIST)~\cite{sank2016measurement,khezri2023measurement}. In the experiment of Ref.~\cite{kurilovich2025high}, the authors used a transmon qubit with $\omega_q/2\pi = 0.76$ GHz and a readout resonator with $\omega_r/2\pi = 9.2$ GHz, resulting in a large ratio of $\omega_r/\omega_q \approx 12$. This choice of parameters enabled fast and high-fidelity readout using $\bar{n}$ approaching 100 photons. Since sub-gigahertz transmon qubit frequencies introduce additional issues, one could imagine scaling up all frequencies in this experiment, such that the qubit frequency is within the more typical range of 3-6 GHz. However, doing so while maintaining a ratio $\omega_r/\omega_q$ above 10 would require a readout resonator with a frequency in the range of 30-60 GHz, i.e., approaching or exceeding the superconducting gap. 

\begin{figure}[t]
    \centering    
    \includegraphics[width=\linewidth]{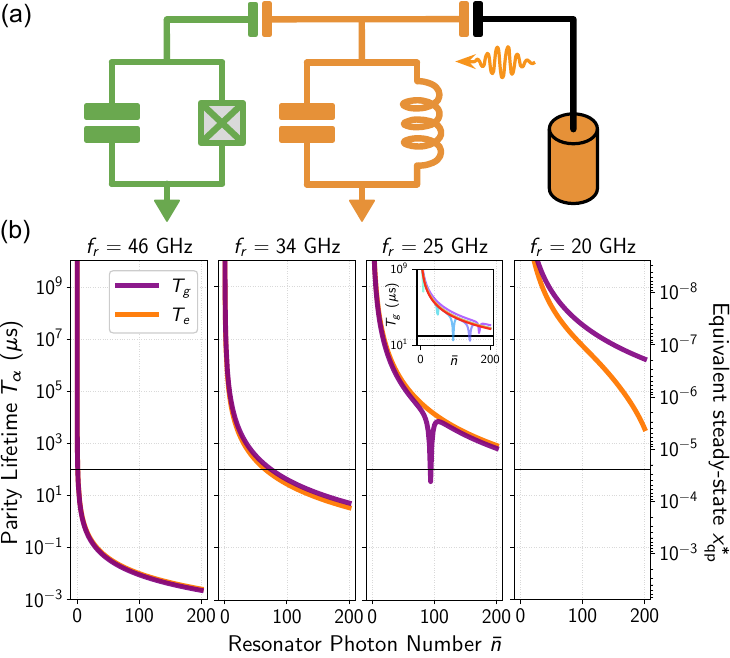}
    \caption{\textbf{Pair-breaking in high-frequency readout.} (a) We consider a readout resonator of frequency $\omega_r$ coupled to a transmon qubit with $E_J/h = 12.85$ GHz, $E_C/h = 218$ MHz, and $n_g = 0.1$, such that the bare qubit frequency is $\omega_q/2\pi = 4.5$ GHz. (b) We plot charge-parity switching lifetimes $T_g$ and $T_e$ starting from the ground and excited states of the qubit, respectively, as a function of resonator photon number $\bar{n}$, for $f_r = \omega_r/2\pi \in \{46, 34, 25, 20\}$ GHz. At each value of $\omega_r$, we adjust the qubit-resonator coupling $g$ so as to maintain a dispersive shift $\chi/2\pi = 1$ MHz. The horizontal black lines indicate a parity lifetime of 100 $\mu$s, and the secondary axis converts the parity lifetimes to an equivalent steady-state QP fraction, $x_{\rm qp}^\ast$ (see the text). In the inset, we plot $T_g$ versus $\bar{n}$ at $f_r = 25$ GHz for several values of the gate charge $n_g \in [0, 0.5]$.}
    \label{fig:3_readout}
\end{figure}

As shown in Fig.~\ref{fig:3_readout}(a), we consider a readout resonator capacitively coupled to a transmon with $E_J/h = 12.85$ GHz and $E_C/h = 218$ MHz, such that the undriven qubit frequency is $\omega_q/2\pi = 4.5$ GHz. We consider four different resonator frequencies: $\omega_r/2\pi \in \{46, 34, 25, 20\}$ GHz. At each value of $\omega_r$, we adjust the qubit-readout coupling $g$ so as to maintain a resonator dispersive shift $\chi/2\pi = 1$ MHz \footnote{Expressions for the dispersive shift in terms of the qubit and resonator parameters in the detuned regime can be found in Ref.~\cite{kurilovich2025high}. For completeness, the specific expression that we use in our numerical simulations is given by: $\hbar\chi = -8E_C (\hbar g)^2(\hbar\omega_r)^2 /[(\hbar\omega_r)^2 -E_{\rm ge}^2][(\hbar\omega_r)^2 -E_{\rm ef}^2]$, where $E_{\alpha\beta} = E_{\beta} - E_{\alpha}$ are bare qubit transition energies.}. In Fig.~\ref{fig:3_readout}(b), we plot the charge-parity lifetimes $T_g$ and $T_e$ versus photon number $\bar{n}$ and find that the lifetimes depend strongly on the resonator frequency. The four chosen frequencies require two, three, four, and five photon absorption processes to generate QPs, respectively, and thus a larger value of $\bar{n}$ is required to achieve the same level of pair-breaking as $\omega_r$ is reduced. In the main panels of Fig.~\ref{fig:3_readout}(b), we have set the transmon offset charge $n_g$ to a nonzero value of $n_g = 0.1$, such that no transitions are parity forbidden. We also convert the QP generation rates $\Gamma_\alpha = T_\alpha^{-1}$ into equivalent steady-state QP fractions $x_{\rm qp}^\ast = \sqrt{c_g/c_r}$, where $c_g = \Gamma_\alpha/N_{\rm cp}$ and $c_r$ are the coefficients for QP generation and recombination, respectively \cite{wang2014measurement}. We use typical values based on the literature for the number of Cooper pairs $N_{\rm cp} = 2\times10^6$ and for $c_r = 1/(120\text{ ns})$ \cite{harrington2024synchronous, diamond2022distinguishing}. As depicted in Fig.~\ref{fig:3_readout}(b), millisecond-level parity lifetimes then correspond to QP fractions of $x_{\rm qp}^\ast \sim 10^{-5}$. These generated QPs have the potential to tunnel across the JJ again, thus causing additional qubit errors~\cite{diamond2022distinguishing}.

At the resonator frequency $\omega_r/2\pi = 25$ GHz, we observe a feature in $T_g$ at approximately $\bar{n} \approx 100$ photons. This corresponds to a conventional MIST process, i.e., a resonance in the Floquet quasi-energy spectrum; here, specifically between the driven-transmon ground and seventh excited states. The hybridization between these states enhances the QP generation rate at this point leading to a dip in $T_g$ (since more final states are available for pair-breaking transitions). In the inset, we plot $T_g$ versus $\bar{n}$ for several values of $n_g \in [0, 0.5]$ and observe multiple such multiphoton resonances that lead to sharp drops in the parity lifetime at different photon numbers. The difference between the parity lifetimes $T_g$ and $T_e$ in the plot with $\omega_r/2\pi = 20$ GHz can likewise be attributed to similar multiphoton resonances. This interplay between conventional MIST and photon-assisted QP generation arises because both processes rely on the nonlinearity of the JJ; however, while MIST is suppressed at higher readout frequencies (given a fixed transmon qubit frequency), the QP generation process is enhanced. We conclude that there exists a trade-off in high-frequency readout schemes between (\textit{i}) increasing the ratio $\omega_r/\omega_q$ to reduce the onset of MIST and (\textit{ii}) keeping $\omega_r$ low enough to prevent excessive QP generation.

\section{Flux-Coupled Drives\label{sec:4_Flux_Coupled}}
\subsection{Extension to time-dependent external flux \label{subsec:4A-flux-drive}}
We now extend the theoretical model developed in Sec. \ref{sec:2_Theoretical_Model} to the case of flux-coupled drives. To this end, we consider a flux-tunable transmon with its JJ replaced by a superconducting quantum interference device (SQUID) loop, threaded by an external flux $\Phi_{\rm e}$ [see Fig.~\ref{fig:4_floquet0pi}(a)]. The Hamiltonian for this SQUID transmon is \cite{koch2007transmon, krantz2019quantum, blais2021circuit}:
\begin{equation}
    \hat{H}_q = 4E_C(\hat{n} - n_g)^2 - E_{J1}\cos(\hat{\varphi}_1) - E_{J2}\cos(\hat{\varphi}_2).
    \label{eq:H_SQUID_transmon}
\end{equation}
As before, $E_C$ is the charging energy, $E_{J1}$ and $E_{J2}$ are the Josephson energies of the two junctions in the SQUID, and $n_g$ is the offset charge. The branch phases $\hat{\varphi}_{1/2}$ are constrained by the fluxoid quantization condition~\cite{barone1982physics, orlando1991foundations, tinkham2004introduction}
\begin{equation}
    \hat{\varphi}_1 - \hat{\varphi}_2 = \varphi_{\rm e}+2\pi k,
\end{equation}
for integer $k$, where $\varphi_{\rm e} \equiv  2\pi\Phi_{\rm e}/\Phi_0$ is the reduced external flux, and $\Phi_0$ is the superconducting flux quantum. This constraint gives rise to just a single effective degree of freedom $\hat{\varphi}$ for the circuit. When the external flux depends on time, the branch phases $\hat{\varphi}_{1/2}$ will also be time-dependent and can be written in terms of $\hat{\varphi}$ as~\cite{you2019timedepflux}
\begin{align}
    \begin{split}
        \hat{\varphi}_1(t) &= \hat{\varphi} + c_1\varphi_{\rm e}(t), \\ 
        \hat{\varphi}_2(t) &= \hat{\varphi} - c_2\varphi_{\rm e}(t).
    \end{split}
    \label{eq:time_dep_flux_branches}
\end{align}
 Here, $c_1$ and $c_2$ are device-specific coefficients that satisfy $c_1 + c_2 = 1$; they represent the allocation of external flux between the two junctions of the SQUID loop and depend on the geometry of both the device and the bias circuitry. Note that we take $\hat{\varphi}$, $\hat{\varphi}_1$, and $\hat{\varphi}_2$ to be compact variables here and thus fix $k = 0$ \cite{devoret2021circle}.

In multijunction superconducting circuits, photon assisted QP generation and tunneling can occur across each of the constituent JJs~\cite{catelani2011relaxation, diamond2022distinguishing}. Therefore, for the SQUID transmon, Eq.~\eqref{eq:1_general_qubit_QP_coupling} becomes
\begin{equation}
    \hat{H}(t) = \hat{H}_q(t) + \hat{H}_{\rm QP} + \sum_{j} \hat{H}_T^j(t),
\end{equation}
where $j \in \{1, 2\}$ indexes the two junctions of the SQUID loop. Here, $\hat{H}_q(t)$ for the SQUID transmon is obtained by substituting $\hat{\varphi}_1(t)$ and $\hat{\varphi}_2(t)$ [Eq.~\eqref{eq:time_dep_flux_branches}] into Eq.~\eqref{eq:H_SQUID_transmon}. The QP Hamiltonian $\hat{H}_{\rm QP}$ is the same as in Eq.~\eqref{eq:H_QP_bath}, and the tunneling Hamiltonians $\hat{H}_T^j(t)$ for the two junctions are \footnote{Here, as in Eqs. (\ref{eq:H_T_electron_tunneling}--\ref{eq:H_T_QP_and_PAT}), we can drop any double-counting correction terms when setting up the Hamiltonian, since they will not be relevant for Fermi's golden rule.}
\begin{align}
\begin{split}
    \hat{H}_T^j(t) &= \tau_j \sum_{l, r, s} \sigma(s) \bigg[\left(u_r v_l+v_r u_l\right) \cos \bigg(\frac{\hat{\varphi}_j(t)}{2}\bigg)\\ &+i\left(u_r v_l-v_r u_l\right) \sin \bigg(\frac{\hat{\varphi}_j(t)}{2}\bigg)\bigg]\hat{\gamma}_{r s}^{\dagger} \hat{\gamma}_{l \bar{s}}^{\dagger} + {\rm h.c.}
\end{split}
\label{eq:H_T_multi_junction}
\end{align}
From here, the derivation follows the procedure described in Sec.~\ref{sec:2_Theoretical_Model}, resulting in Floquet extended-space rates for each junction: $\Gamma_{ \alpha\beta, j}^{(n)}$. We then sum over these rates for the two junctions in the SQUID to get the total transition rate between Floquet extended states $\ket{\tilde{\Phi}_{\alpha, 0}}$ and $\ket{\tilde{\Phi}_{\beta, -n}}$: 
\begin{equation}
    \Gamma_{\alpha\beta}^{(n)} = \Gamma_{ \alpha\beta, 1}^{(n)} + \Gamma_{ \alpha\beta, 2}^{(n)}.
\end{equation}
Finally, as before, we project back onto the qubit Hilbert space by summing over the Floquet photon-number index: $\Gamma_{\alpha\beta} = \sum_n \Gamma_{\alpha\beta}^{(n)}$. As a last remark, we note that it would be straightforward to further generalize the results presented here to many-junction circuits, following Refs.~\cite{catelani2011relaxation} and~\cite{you2019timedepflux}. This could be applied, for example, to calculate drive-induced pair-breaking across the array junctions of a fluxonium qubit. 

\subsection{Strongly-driven Floquet qubits}

Floquet engineering in superconducting qubits involves using a strong modulation drive to implement an effective potential or interaction beyond what is possible via static circuit elements alone; for instance, to realize a protected qubit \cite{mundada2020floquet,gandon2022floquet, nguyen2024programmable, thibodeau2024floquet, geier2024self, kapitzonium2024,lewellen2025frozonium, sellem2025dissipative}. One such example involves using a flux-modulated symmetric SQUID to realize a Floquet $0-\pi$ qubit with an effective $\cos(2\hat{\varphi})$ potential arising from the drive \cite{kapitzonium2024}. We consider this qubit as a case study for calculating pair-breaking rates under strong flux-coupled microwave drives. 

The Hamiltonian for a symmetric SQUID is given by Eq.~\eqref{eq:H_SQUID_transmon} with $E_{J1} = E_{J2} = E_J$:
\begin{align}
    \begin{split}
        \hat{H}_q(t) &= 4E_C(\hat{n} - n_g)^2 - E_J\cos\Big(\hat{\varphi} + \frac{1}{2}\varphi_{\rm e}(t)\Big) \\ &\qquad - E_J\cos\Big(\hat{\varphi} - \frac{1}{2}\varphi_{\rm e}(t)\Big) \\
        &= 4E_C(\hat{n} - n_g)^2 - 2E_J\cos\bigg(\frac{\varphi_{\rm e}(t)}{2}\bigg)\cos(\hat{\varphi}).
    \end{split}
    \label{eq:H_symmetric_SQUID}
\end{align}
In addition to identical junctions, we have assumed that the external flux bias circuitry has an equal and opposite coupling to the two junctions of the SQUID. We consider an external flux with both dc and ac components: $\varphi_{\rm e}(t) = \varphi_{\rm dc} + \varphi_{\rm ac}\sin(\omega_d t)$. For the Floquet $0-\pi$ qubit, we set $\varphi_{\rm dc} = 0$ and keep only the sinusoidal modulation. When the driving frequency $\omega_d$ is large compared with the intrinsic energy scale of the static circuit, we can approximate the dressed low-energy spectrum of the driven qubit using a static effective Hamiltonian expressed as a perturbative expansion in  $1/\omega_d$ \cite{kapitzonium2024, rahav2003, jaya2022static} (see Appendix~\ref{app:B_Floquet_Theory}). To second order, the effective Hamiltonian that arises from Eq.~\eqref{eq:H_symmetric_SQUID} is given by
\begin{align}
    \begin{split}
        \hat{H}_{\rm eff} &= 4E_C(\hat{n} - n_g)^2 - 2E_J J_0\Big(\frac{\varphi_{\rm ac}}{2}\Big)\cos(\hat{\varphi}) \\ &\qquad -\frac{4E_C E_J^2}{\omega_d^2}\sum_{n = 1}^\infty \bigg[\frac{J_{2n}(\varphi_{\rm ac}/2)}{n}\bigg]^2 \cos(2\hat{\varphi}),
    \end{split}
    \label{eq:H_eff_kapitza}
\end{align}

\begin{figure}[t]
    \centering    
    \includegraphics[width=\linewidth]{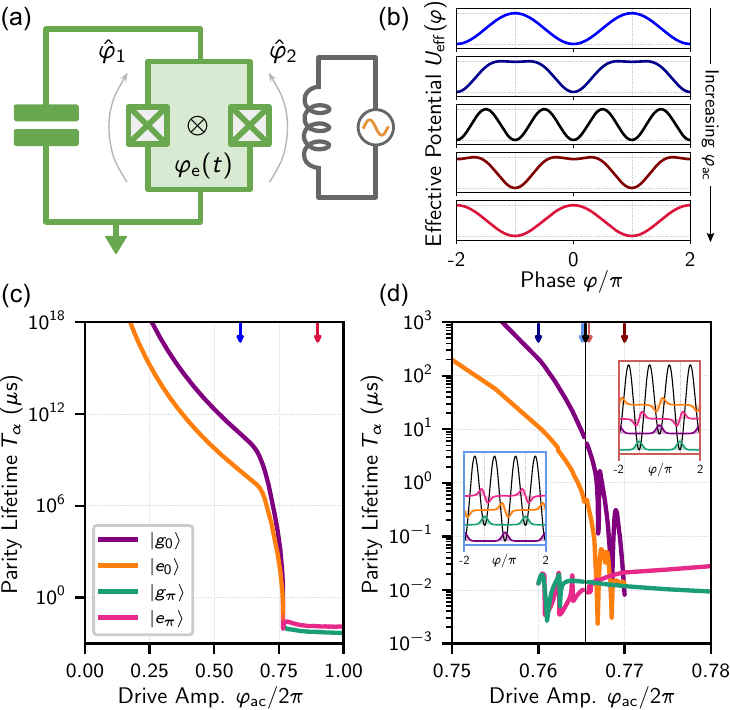}
    \caption{\textbf{Pair-breaking in a flux-driven Floquet qubit.} (a) Circuit schematic for a symmetric SQUID threaded by an external flux $\varphi_{\rm e}(t) = \varphi_{\rm ac}\sin(\omega_d t)$ [Eq.~\eqref{eq:H_symmetric_SQUID}]. Here, QP generation can occur across each of the junctions in the SQUID. (b) Top to bottom: Effective potentials $U_{\rm eff}$ [cf. Eq. \eqref{eq:H_eff_kapitza}] at different drive amplitudes $\varphi_{\rm ac}/2\pi \in [0.5, 0.76, \varphi_{\rm ac}^\star/2\pi, 0.77, 0.9]$, where $\varphi_{\rm ac}^\star/2\pi \approx 0.76547$ is the drive amplitude that eliminates the $\cos(\hat{\varphi})$ component of $U_{\rm eff}$, leaving only the $\cos(2\hat{\varphi})$ component. Plotted potentials are normalized for clarity. (c) Parity lifetimes $T_\alpha$ for Floquet states that map onto $\ket{g_0}$, $\ket{e_0}$, $\ket{g_\pi}$, and $\ket{e_\pi}$ (ground and excited states in the wells at $\varphi=0$ and $\pi$, respectively). See the main text for circuit parameters. (d) Enlarged view of the data from panel (c) exhibiting a crossover at $\varphi_{\rm ac}^\star$ (black solid line), as the system's ground state changes from $\ket{g_0}$ to $\ket{g_\pi}$. The obtained lifetimes for $\ket{g_\pi/e_\pi}$ when $\varphi_{\rm ac} < \varphi_{\rm ac}^\star$ and for $\ket{g_0/e_0}$ when $\varphi_{\rm ac} > \varphi_{\rm ac}^\star$, respectively, exhibit large jumps due to the avoided crossings that these states undergo as the potential is adiabatically deformed. Bound states are present in both wells only for $0.76 \lesssim \varphi_{\rm ac}/2\pi \lesssim 0.77$.  The insets depict the effective potential and phase-basis wavefunctions at flux drive amplitudes $\varphi_{\rm ac} = (1 \pm 0.0005)\varphi_{\rm ac}^\star$.}
    \label{fig:4_floquet0pi}
\end{figure}
\noindent where $J_\nu$ is the $\nu$th Bessel function of the first kind. Since the $\cos(\hat{\varphi})$ term in Eq.~\eqref{eq:H_eff_kapitza} is proportional to $J_0(\varphi_{\rm ac}/2)$, it can be made to vanish at a specific drive amplitude $\varphi_{\rm ac}^\star/2\pi \approx 0.76547$ that corresponds to a zero of the Bessel function. At $\varphi_{\rm ac}^\star$, the resulting $\cos(2\hat{\varphi})$ potential has two degenerate ground states that are localized around $\varphi = 0$ and $\varphi = \pi$, respectively. Then, for flux drive amplitudes exceeding $\varphi_{\rm ac}^\star$, the sign of $J_0$ changes from positive to negative and, thus, the leading term in the effective potential becomes proportional to $+\cos(\hat{\varphi})$ rather than the usual $-\cos(\hat{\varphi})$. We plot this effective potential $U_{\rm eff}$ for several values of the flux drive amplitude $\varphi_{\rm ac}$ in Fig.~\ref{fig:4_floquet0pi}(b). We additionally plot the phase-basis wavefunctions at drive amplitudes smaller than and greater than $\varphi_{\rm ac}^\star$ in the inset of Fig.~\ref{fig:4_floquet0pi}(d) and observe that the overall system ground-state transitions from $\ket{g_0}$ to $\ket{g_\pi}$, where $\ket{g_0}$ and $\ket{g_\pi}$ denote the lowest energy states in the wells at $\varphi = 0$ and $\varphi = \pi$, respectively. 

To numerically simulate pair-breaking in the SQUID circuit above, we promote the Hamiltonian from Eq.~\eqref{eq:H_symmetric_SQUID} to the Floquet extended Hilbert space and calculate transition rates following the procedure outlined in Sec.~\ref{subsec:4A-flux-drive}. At each drive amplitude, we label the obtained extended Floquet states $\ket{\tilde{\Phi}_{\alpha, m}}$ by projecting back onto the original qubit Hilbert space at time $t = 0$, and then identifying states that have the largest overlap with the static eigenstates  of the effective Hamiltonian [Eq.~\eqref{eq:H_eff_kapitza}] (see Appendix~\ref{app:B_Floquet_Theory}). We specifically search for the extended-space states that map onto $\ket{g_0}$, $\ket{e_0}$, $\ket{g_\pi}$, and $\ket{e_\pi}$ (i.e., the ground and excited states in each well), and then use these as the initial states in the Fermi's golden rule calculation. In Fig.~\ref{fig:4_floquet0pi}, we consider a Floquet $0-\pi$ circuit with parameters $E_J/h = 81.6$ GHz, $E_C/h = 10$ MHz, $n_g = 0$, and $\omega_d/2\pi = 10$ GHz. When the drive amplitude is $\varphi_{\rm ac}^\star$, the resulting $\cos(2\hat{\varphi})$ potential thus has a prefactor $E_{J, 2\varphi}/h = 0.5$ GHz, corresponding to a ratio $E_{J, 2\varphi}/E_C = 50$. Although the flux modulation frequency is chosen to be much smaller than $2\Delta/h = 90$ GHz, $\hat{H}_q(t)$ in Eq.~\eqref{eq:H_symmetric_SQUID} still contains several high-frequency harmonics, which can be seen by expanding $\cos(\varphi_{\rm ac}\sin(\omega_dt)/2)$ via the Jacobi-Anger expansion. Each harmonic $k$ can mediate $\lceil2\Delta/k\hbar\omega_d\rceil$-photon pair-breaking transitions.

In Fig.~\ref{fig:4_floquet0pi}(c), we plot the parity lifetimes $T_\alpha$ versus drive amplitude $\varphi_{\rm ac}$, and observe a sharp drop in the lifetimes approaching $\varphi_{\rm ac}^\star$ (as the drive starts to significantly affect the effective potential).  Furthermore, the QP generation rates remain high for flux drive amplitudes $\varphi_{\rm ac} > \varphi_{\rm ac}^\star$, as the ground and excited states shift to being localized around $\varphi = \pi$. We enlarge the region around $\varphi_{\rm ac}^\star$ in Fig.~\ref{fig:4_floquet0pi}(d), and observe that the pair-breaking transition rates for $\ket{g_\pi}$ and $\ket{e_\pi}$ consistently exceed those for $\ket{g_0}$ and $\ket{e_0}$. We hypothesize that this behavior is due to the fact that $\varphi = \pi$ is an unstable equilibrium of an undriven JJ, and thus forcing the eigenstates to be localized at $\varphi = \pi$ [i.e., inverting the effective potential from $-\cos(\hat{\varphi})$ to $+\cos(\hat{\varphi})$] via a strong modulation drive is inherently unstable. This is reflected in the dressed matrix elements of $\cos[\hat{\varphi}/2 \pm \varphi_{\rm ac}\sin(\hat{\vartheta})/4]$ and $\sin[\hat{\varphi}/2 \pm \varphi_{\rm ac}\sin(\hat{\vartheta})/4]$ that mediate the pair-breaking transitions [cf. Eq.~\eqref{eq:H_T_multi_junction}]; these matrix elements are consistently larger for the eigenstates localized at $\varphi = \pi$. At exactly $\varphi_{\rm ac} = \varphi_{\rm ac}^\star$ (not shown on the plot), the parity lifetimes are on the order of $T_\alpha \sim 0.1$ $\mu$s. Altogether, these results demonstrate how strong flux modulation can enhance pair-breaking and therefore limit charge-parity lifetimes in a Floquet qubit.

\section{Conclusion and Outlook\label{sec:5_Conclusion}}
We have presented theory to calculate the rates of pair-breaking transitions from multiphoton-induced QP generation in strongly driven superconducting circuits, and have developed numerical simulations applying this theoretical framework to various examples in circuit quantum electrodynamics. We have found that the rates of such pair-breaking transitions are negligible for typical qubit/readout frequencies and weak drive amplitudes; however, for higher frequency microwave tones or large drive amplitudes, such processes become relevant. Remarkably, photon-assisted QP generation can occur even if the microwave frequency is well below twice the superconducting gap, i.e., $\omega_d \ll 2\Delta/\hbar$. This reflects an induced subgap QP current across the JJ, even in the absence of any pre-existing thermal QPs. While past studies of photon-assisted effects in JJs (e.g., in the context of superconducting mixers and detectors) rely on semi-classical models such as Tien-Gordon theory \cite{tien1963gordon} to describe averaged JJ $I$-$V$ curves \cite{tucker1979quantum, tucker1985feldman, richards1989superconducting, vanduzer1999principles}, our approach treats the qubit degree of freedom fully quantum mechanically and also includes the BCS coherence factors for the QPs. We therefore generalize the results of Ref.~\cite{houzet2019photon} from single-photon absorption with $\hbar\omega_d > 2\Delta$ to multiphoton absorption more broadly. 

In the context of modern circuit quantum electrodynamics, the theoretical framework developed here will be particularly important in the design of next-generation Floquet qubits that typically require strong always-on microwave drives. This type of analysis may also be relevant for Schr\"odinger cat qubits, and therefore complementary to other perturbative analytic approaches used for modeling QP-induced decoherence, such as in Ref. \cite{dubovitskii2024theory}. The Floquet extended space techniques and numerics developed here may also be applied to theoretically model other decoherence channels in microwave-driven qubits. Drive-induced QP generation is additionally an important consideration for novel dispersive readout architectures involving high-frequency readout resonators. One natural extension of this work would be to investigate possible effects of quantum fluctuations in the resonator, which were neglected in our analysis in Sec.~\ref{subsec:detuned_readout}.

Finally, recent advances in gap engineering have significantly reduced the impact of QP tunneling on the coherence of superconducting transmon qubits, in particular through use of different gaps $\Delta_L$ and $\Delta_R$ on the two sides of the junction \cite{diamond2022distinguishing, harrington2024synchronous, mcewen2024resisting}. The photon-assisted QP generation process we investigated here is not suppressed by this asymmetry $\delta\Delta$ but instead only depends on the total gap $2\bar{\Delta}$, where $\bar{\Delta} = (\Delta_L + \Delta_R)/2$ and $\delta\Delta = \Delta_L - \Delta_R$. However, as seen in our work, the rate of pair-breaking transitions is exponentially sensitive to the ratio $2\bar{\Delta}/\hbar\omega_d$. This suggests that photon-assisted QP generation can be suppressed by using thinner films to increase the superconducting gap in aluminum, or by using higher gap materials such as tantalum or niobium~\cite{anferov2024improved}. 

\textit{Note}---During the preparation of this manuscript, we became aware of Ref.~\cite{vlad2025qps}, which independently analyzes QP-induced decoherence in transmon qubits subjected to microwave drives using a perturbative analytic approach. We have verified that our results are in agreement. 

\textbf{Acknowledgements.} We would like to thank Vlad Kurilovich, Pavel Kurilovich, Mykola Kishmar, and Manuel Houzet for insightful discussions related to driven QPs. We also gratefully acknowledge Xu Xiao and Agustin Di Paolo for helpful discussions about Floquet theory and its numerical implementation. This research was funded in part by the U.S. Army Research Office under Award No. W911NF-23-1-0045 and in part by the U.S. Department of Energy, Office of Science, National Quantum Information Science Research Centers, Co-design Center for Quantum Advantage (C2QA) under Contract No. DE-SC0012704; in part by the AWS Center for Quantum Computing; and in part under Air Force Contract No. FA8702-15-D-0001.  S.C. and S.R.J acknowledge support from the NSF Graduate Research Fellowship. M.H. is supported by an appointment to the Intelligence Community Postdoctoral Research Fellowship Program at the Massachusetts Institute of Technology administered by Oak Ridge Institute for Science and Education (ORISE) through an interagency agreement between the U.S. Department of Energy and the Office of the Director of National Intelligence (ODNI). The views and conclusions contained herein are those of the authors and should not be interpreted as necessarily representing the official policies or endorsements, either expressed or implied, of the U.S. Government.

\appendix
\section{BCS HAMILTONIAN \label{app:A_BCS_Setup}}
In this appendix, we give a sketch of the derivation of Eq. \eqref{eq:1_general_qubit_QP_coupling} from BCS theory, and in particular highlight why it is important to develop the framework from Sec.~\ref{sec:2_Theoretical_Model} in a charge-displaced frame. A more complete version of this derivation can be found in Ref.~\cite{supp}.

The Hamiltonian for a BCS superconductor is given by
\begin{equation}
    \hat{H} = \sum_{k,s} \xi_{k} \hat{c}_{ks}^{\dagger} \hat{c}_{ks} + \sum_{k} \big(\tilde{\Delta}\hat{c}_{k\uparrow}^{\dagger} \hat{c}_{-k\downarrow}^{\dagger} + \tilde{\Delta}^\ast\hat{c}_{-k\downarrow} \hat{c}_{k\uparrow} \big),
\end{equation}
where $\hat{c}_{ks}$ is the annihilation operator for an electron with momentum $k$ and spin $s$, $\xi_{k}$ is the kinetic energy of the electron, and $\tilde{\Delta} = \Delta e^{i\phi}$ is the pair potential, which here is a complex parameter that depends on the superconducting gap $\Delta$ and the phase order parameter $\phi$ of the superconductor~\cite{tinkham2004introduction}. If we have two such BCS superconductors with different phases $\phi_{L}$ and $\phi_{R}$, the Hamiltonian for the combined system can be modeled via
\begin{equation}
    \hat{H} = \hat{H}_{L}(\phi_L) + \hat{H}_{R}(\phi_R) + E_C(\hat{N}_L - \mathcal{N}_g)^2 + \hat{H}_T,
    \label{eq:H_BCS_LR}
\end{equation} 
where $\hat{H}_{L/R}$ are the BCS Hamiltonians for the respective superconductors and $\hat{H}_T = \tau\sum_{l, r, s}  [\hat{c}_{rs}^\dagger \hat{c}_{ls} + {\rm h.c.}]$ is the tunneling Hamiltonian~\cite{grabert1992single}. When the tunneling amplitude $\tau$ is weak, Eq.~\eqref{eq:H_BCS_LR} models a Josephson tunnel junction (JJ). We have also included an electrostatic energy term for electrons on one of the islands (without loss of generality, the left island here). This term depends on the single-electron charging energy $E_C = e^2/2(C_J + C_g)$, the electron dc offset charge $\mathcal{N}_g = C_gV_g/e$, and the number of electrons $\hat{N}_L = \sum_{l, s} \hat{c}_{ls}^{\dagger} \hat{c}_{ls}$ in the left superconducting island \cite{lafarge1993measurement, esteve1994measurement, falci1994tunneling, neumann1995charge}. Here, $C_J$ and $C_g$ are the intrinsic JJ capacitance and the capacitance between the island and a gate electrode with gate voltage $V_g$, respectively.

Let us now suppose that the right lead is grounded and we have an oscillating voltage $V(t)$ across the JJ. In this case, we must modify the BCS Hamiltonian of the left lead to $\hat{H}_{L}(\phi_L) \to \hat{H}_{L}\big(\phi_L + \int dt\, 2eV(t)/\hbar\big) - eV(t)\hat{N}_L$, to be consistent with the second Josephson relation relating the phase difference between the leads to the voltage across the JJ. The source of this oscillating voltage $V(t)$ is irrelevant here: it could come from a direct voltage bias, irradiation by microwaves, or the voltage induced by a time-dependent gate offset $V_g(t)$ that results in $V(t) = C_gV_g(t)/(C_J + C_g)$. In all cases, the induced voltage couples to the superconducting phase variable as well as to the electrons themselves via $\hat{N}_L$. 

Solving for the BCS ground state of $\hat{H}_L + \hat{H}_R$ is difficult when $\hat{H}_L$ depends on time. However, we may simplify the Hamiltonian by performing two unitary transformations. The first transforms the electronic annihilation operators $\hat{c}_{ls} \rightarrow e^{i\phi_L/2} \hat{c}_{ls}$ and $\hat{c}_{rs} \rightarrow e^{i\phi_R/2} \hat{c}_{rs}$ in the two leads. The tunneling Hamiltonian after making this transformation becomes
\begin{equation}
    \hat{H}_T = \tau \sum\limits_{l, r, s} \big[e^{i\varphi/2} \hat{c}_{rs}^\dagger \hat{c}_{ls} + {\rm h.c.}\big],
    \label{eq:HT_glazman}
\end{equation}
where $\varphi = \phi_L - \phi_R$ is the phase difference. The second transformation is a time-dependent unitary $\hat{U}(t) = \exp[i\int eV(t)\hat{N}_L/\hbar dt]$ which both removes the time-dependent phase of the left lead and cancels out the term $-eV(t)\hat{N}_L$. The resulting transformed BCS Hamiltonians will be time-independent and have zero phase \cite{glazman2021bogoliubov}. Therefore, by performing the Bogoliubov transformation from Sec.~\ref{sec:2_Theoretical_Model} on $\hat{H}_L + \hat{H}_R \equiv \hat{H}_{\rm QP}$, we arrive at Eq.~\eqref{eq:H_QP_bath}. At the same time, $\hat{H}_T$ from Eq.~\eqref{eq:HT_glazman} transforms into Eq.~\eqref{eq:H_T_electron_tunneling}. 

The only remaining step is to define the superconducting qubit degree of freedom, which involves promoting $\varphi$ to an operator $\hat{\varphi}$. Formally, this is done by performing second-order perturbation theory on $\hat{H}_L + \hat{H}_R$ and considering $\hat{H}_T$ as the perturbation \cite{martinis2004superconducting,girvin2019modern,liao2024circuit}. The resulting effective low-energy qubit Hamiltonian is exactly Eq.~\eqref{eq:H_transmon_disp_frame}, and the total qubit-QP Hamiltonian is given by Eq.~\eqref{eq:1_general_qubit_QP_coupling}. In obtaining this effective Hamiltonian, we can also transform from electronic operators to Cooper pair operators and replace the electronic charging energy term above by an equivalent Cooper pair charging energy term $4E_C(\hat{n} - n_g)^2$, where $\hat{n}$ represents the number of Cooper pairs that have tunneled across the JJ. We also absorb the average electron background charge into the Cooper pair offset charge $n_g$, which is otherwise related to $\mathcal{N}_g$ by a factor of 1/2 (see Ref.~\cite{supp} for more details).

As a final remark, we note that the conventional Hamiltonian for a driven transmon (or a Cooper-pair box) is often written as Eq.~\eqref{eq:H_transmon_Rabi_frame}, with the time-dependence of the drive entering via a linear ``charge drive'' term $2eV(t)\hat{n}$. By contrast, here we incorporate the drive into the qubit Hamiltonian via a time-dependent phase as in Eq.~\eqref{eq:H_transmon_disp_frame}. While either is correct if we trace out the QP degrees of freedom and consider only the qubit dynamics, we find that Eq.~\eqref{eq:H_transmon_disp_frame} is a better starting point for the Hamiltonian of a driven transmon when modeling QP-induced losses, since it directly corresponds to the qubit term in the full qubit-QP model when correctly accounting for the time-dependence of the QP bath  [Eq.~\eqref{eq:1_general_qubit_QP_coupling}]. The transformation between Eqs. \eqref{eq:H_transmon_disp_frame} and \eqref{eq:H_transmon_Rabi_frame} is given in Sec.~\ref{subsec:drive_qubits}.

\section{FLOQUET THEORY \label{app:B_Floquet_Theory}}
\subsection{Overview}

Floquet theory is a powerful framework for analyzing periodically driven systems. In this appendix, we provide a self-contained introduction to the extended Hilbert space Floquet formalism used throughout this work \cite{shirley1965solution, sambe1973steady, drese1999floquet, grifoni1998driven, eckardt2015high, di2022extensible, rudner2020floquet, kitagawa2011transport}. 

We begin by considering a generic periodically driven system with a Hamiltonian $\hat{H}$ satisfying $\hat{H}(t) = \hat{H}(t + \mathbb{T})$ with period $\mathbb{T}$. Equivalently, we can define the frequency $\omega_d = 2\pi/\mathbb{T}$ of the drive. The dynamics of this system are governed by the Schr{\"o}dinger equation
\begin{equation}
    i\hbar\frac{d}{dt}\ket{\psi(t)} = \hat{H}(t)\ket{\psi(t)}.
    \label{eq:time_dep_schrodinger}
\end{equation}
When the Hamiltonian is periodic in time, the eigenstates $\ket{\psi_\alpha(t)}$ of the system take the form
\begin{equation}
    \ket{\psi_\alpha(t)} = e^{-i\epsilon_\alpha t/\hbar}\ket{\phi_\alpha(t)}, \quad \ket{\phi_\alpha(t)} = \ket{\phi_\alpha(t + \mathbb{T})},
    \label{eq:floquet_modes}
\end{equation}
according to Floquet's theorem~\cite{floquet1883equations}. Here $\alpha$ indexes the qubit state, and we have so-called Floquet quasi-energies $\epsilon_\alpha$ which are only defined modulo $\hbar\omega_d$ here. The states $\ket{\psi_\alpha(t)}$ are written in terms of the Floquet modes $\ket{\phi_\alpha(t)}$, which are also time-periodic with period $\mathbb{T}$. Given this fact, we can then further decompose each Floquet mode as a Fourier series
\begin{equation}
    \ket{\phi_\alpha(t)} = \sum_{m} e^{im\omega_d t}\ket{\phi_{\alpha}^{(m)}},
    \label{eq:floquet_modes_fourier}
\end{equation}
where the summation index runs over all integers $m \in \Z$. Our goal is now to find an equation to obtain the Fourier components $\ket{\phi_{\alpha}^{(m)}}$, as well as the corresponding Floquet quasi-energies. We do so by first plugging Eq.~\eqref{eq:floquet_modes} into Eq.~\eqref{eq:time_dep_schrodinger} to get a Schr{\"o}dinger equation for the Floquet modes:
\begin{equation}
    \bigg[\hat{H}(t) - i\hbar\frac{d}{dt}\bigg]\ket{\phi_\alpha(t)} = \epsilon_\alpha\ket{\phi_\alpha(t)}. 
    \label{eq:floquet_mode_schrodinger}
\end{equation}
The operator in parentheses is a precursor to the Floquet Hamiltonian, here expressed in the original qubit Hilbert space. If we now substitute Eq.~\eqref{eq:floquet_modes_fourier} into Eq.~\eqref{eq:floquet_mode_schrodinger} and also expand $\hat{H}(t) = \sum_m e^{im\omega_d t}\hat{H}^{(m)}$ as a Fourier series, we get
\begin{equation}
    (\epsilon_\alpha - m\hbar\omega_d)\ket{\phi_{\alpha}^{(m)}} = \sum_n \hat{H}^{(m - n)}\ket{\phi_{\alpha}^{(n)}}.
    \label{eq:floquet_matrix_eq}
\end{equation}
At this point, one can solve Eq.~\eqref{eq:floquet_matrix_eq} as a matrix equation by stacking up all of the Fourier components $\{\ket{\phi_{\alpha}^{(m)}}\}$ into a vector $\ket{\lambda_\alpha}$ which satisfies $\hat{H}^F\ket{\lambda_\alpha} = \epsilon_\alpha \ket{\lambda_\alpha}$, where the Floquet Hamiltonian is an infinite-dimensional block matrix of the form
\begin{equation}
    \hat{H}^F = \begin{bmatrix}
        \ddots & \vdots & \vdots & \vdots & \iddots \\
        \hdots & \hat{H}_0 - \hbar\omega_d & \hat{H}_{-1}  & \hat{H}_{-2} & \hdots \\
        \hdots & \hat{H}_{1} & \hat{H}_0 & \hat{H}_{-1} & \hdots \\
        \hdots & \hat{H}_{2} & \hat{H}_{1} & \hat{H}_0 + \hbar\omega_d & \hdots \\
        \iddots & \vdots & \vdots & \vdots & \ddots
    \end{bmatrix}.
    \label{eq:HF_matrix}
\end{equation}
Note that $\hat{H}^F$ can be diagonalized numerically to great accuracy by truncating the full infinite-dimensional matrix to a finite size $d \times (2M_{\rm max} + 1)$, where $d$ is the (possibly truncated) dimension of the qubit Hilbert space and $M_{\rm max}$ is the largest Fourier component considered. When performing numerical diagonalizations, it is important to verify the convergence of the results with $M_{\rm max}$. We have carefully checked this for all of the numerical simulations presented in this work (see Appendix~\ref{sec:sim_details} for additional details about how these numerical simulations were set up). 

Returning to our goal of obtaining the Floquet modes, one can reconstruct $\ket{\phi_\alpha(t)}$ using the Fourier coefficients $\ket{\phi_\alpha^{(m)}}$ by applying the projector $\hat{P}(\omega_d t)$; this is defined as the $d \times [d \times (2M_{\rm max} + 1)]$-dimensional rectangular matrix $\hat{P}(\omega_d t) = [\ldots,\,\, e^{im\omega_d t}\hat{I}_{d}, \,\, e^{i(m+1)\omega_d t}\hat{I}_{d}, \,\,\ldots]$, where $\hat{I}_d$ is the $d$-dimensional qubit identity matrix. We then compute the Floquet mode at any time $t$ from a numerically obtained eigenvector $\ket{\lambda_\alpha}$ via $\ket{\phi_\alpha(t)} = \hat{P}(\omega_d t)\ket{\lambda_\alpha}$ \cite{rudner2020floquet}.

Finally, we remark that when the driving frequency $\omega_d$ is much larger than the intrinsic energy scales of $\hat{H}_0$, it is possible to block-diagonalize $\hat{H}^F$ such that the diagonal blocks are all of the form $\hat{H}_{\rm eff} + m\hbar\omega_d$ with $m \in \Z$. This defines the effective Hamiltonian $\hat{H}_{\rm eff}$, which acts on the qubit Hilbert space and approximately captures the effect of the drive on the qubit. The effective Hamiltonian $\hat{H}_{\rm eff}$ is typically constructed perturbatively using a power series in the small parameter $1/\omega_d$. Explicit expressions for this construction can be found in Refs.~\cite{jaya2022static, rahav2003, eckardt2015high}. 

\subsection{Coupled-mode extended Hilbert space picture}

We now present an alternative perspective on the construction of the Floquet Hamiltonian $\hat{H}^F$ that we believe provides more intuition for the Floquet problem and its numerical implementation. 

For concreteness, we first outline this approach for the specific example of a monochromatic charge-driven qubit with Hamiltonian $\hat{H}(t) = \hat{H}_q + \hbar\Omega\hat{n}\cos(\omega_d t)$ [cf. Eq. \eqref{eq:H_transmon_Rabi_frame}], and later generalize to arbitrary periodic driving. Here $\hat{H}_q$ is the undriven qubit Hamiltonian with eigenenergies and eigenstates labeled $E_\alpha$ and $\ket{\alpha}$, respectively. To construct the Floquet Hamiltonian from $\hat{H}(t)$, we first define the drive phase $\theta(t) = \omega_d t$. We can then promote $\theta(t)$ to a $2\pi$-periodic quantum degree of freedom $\hat{\vartheta}$ with conjugate variable $\hat{m}$ satisfying $[\hat{\vartheta}, \hat{m}] = i$. Since $\hat{\vartheta}$ is $2\pi$-periodic, the operator $\hat{m}$ will be quantized to integer values $m \in \mathbb{Z}$. One can thus write down $\cos(\hat{\vartheta}) \mapsto \frac{1}{2}\sum_m \op{m+1}{m} + {\rm h.c.}$ and $\hat{m} \mapsto \sum_m m\op{m}{m}$ in the $m$-basis. Constructing the drive operators in this way makes it clear how they can be implemented numerically: readers familiar with superconducting circuits might observe that $\hat{m}$ and $\hat{\vartheta}$ have the same structure as the charge and phase operators of a transmon written in its charge basis \cite{shoumik-practical-floquet}. However, $m$ is better interpreted here as a photon number; specifically, the number of drive photons absorbed or emitted. 

The promotion step $\theta(t) \mapsto \hat{\vartheta}$ above effectively defines a Floquet extended Hilbert space $\mathcal{H} \otimes \mathcal{F}$ where $\mathcal{H}$ is the original qubit Hilbert space and $\mathcal{F}$ is the Floquet space spanned by the Floquet photon-number basis states $\ket{m}$. 
Given that the conjugate momentum $\hat{m}$ may be written as $\hat{m} \to -i\partial_\theta$ in the $\theta$ basis, we can make an association between $\hat{H}(t) - i\hbar\partial_t$ and the full Floquet Hamiltonian $\hat{H}^F$, which is written in the extended Hilbert space as
\begin{equation}
    \hat{H}^F = \hat{H}_q + \hbar\Omega\hat{n}\cos(\hat{\vartheta}) + \hbar\omega_d\hat{m}.
    \label{eq:HF_dressed_mode}
\end{equation}
The Floquet Hamiltonian here is exactly what one would arrive at by plugging $\hat{H}(t)$ into Eq.~\eqref{eq:HF_matrix}, which shows the full infinite-dimensional matrix form of $\hat{H}^F$. However, we can now also interpret Eq.~\eqref{eq:HF_dressed_mode} as the Hamiltonian for a two-mode coupled system consisting of the qubit and the drive, with two distinct quantum numbers $\alpha$ and $m$, respectively. We can therefore use static circuit diagonalization libraries (e.g., \texttt{jaxquantum} \cite{jha2024jaxquantum}) to construct $\hat{H}^F$ and solve for its spectrum. When the drive is turned off ($\Omega \to 0$), the Floquet eigenstates $\ket{\Phi_{\alpha, m}}$ of $\hat{H}^F$ as defined in Eq.~\eqref{eq:HF_dressed_mode} will be product states $\ket{\Phi_{\alpha, m}} \equiv \ket{\alpha}\ket{m}$ with energies $E_{\alpha, m} = E_\alpha + m\hbar\omega_d$. For this reason, the Floquet extended Hilbert space is often called the replicas picture; it is also referred to as the Shirley \cite{shirley1965solution} or Sambe \cite{sambe1973steady} space picture. Returning to Eq.~\eqref{eq:HF_dressed_mode}, we note that as the drive amplitude $\Omega$ is increased, the qubit and drive degrees of freedom will hybridize, giving rise to dressed states $\ket{\tilde{\Phi}_{\alpha, m}} \equiv \ket{\widetilde{\alpha, m}}$ with associated energies $\tilde{E}_{\alpha, m}$. These dressed energies are related to the original Floquet quasi-energies $\epsilon_\alpha$ via $\epsilon_\alpha = \tilde{E}_{\alpha, m} \mod \hbar\omega_d$.

To generalize this extended Hilbert space (i.e., coupled-mode) approach to arbitrary periodic driving, we expand $\hat{H}(t)$ as a Fourier series $\hat{H}(t) = \sum_k e^{ik\omega_d t}\hat{H}^{(k)}$ as done before. The Floquet Hamiltonian is then given by \footnote{We could also further generalize to multitone drives with incommensurate frequencies $\omega_1$ and $\omega_2$. Here, the Floquet Hamiltonian would be $\hat{H}^F = \hat{H} + \hbar\omega_1\hat{m}_1 + \hbar\omega_2\hat{m}_2$ where $\hat{H}$ is the extended space promotion of $\hat{H}(t)$, and $\hat{m}_1$ and $\hat{m}_2$ are the respective conjugate variables to the drive phases $\theta_1(t) = \omega_1 t$ and $\theta_2(t) = \omega_2 t$.}:
\begin{equation}
    \hat{H}^F = \sum_k \hat{H}^{(k)} e^{ik\hat{\vartheta}} + \hbar\omega_d\hat{m}.
    \label{eq:HF_arbitrary_drive}
\end{equation}
The operators $e^{ik\hat{\vartheta}}$ are expressed in the Floquet photon-number basis using $e^{ik\hat{\vartheta}} \mapsto \sum_m \op{m+k}{m}$, which again makes clear how we can numerically construct the Floquet Hamiltonian [Eq.~\eqref{eq:HF_arbitrary_drive}]. As a concrete example, let us consider the charge-displaced transmon from Eq.~\eqref{eq:H_transmon_disp_frame}, for which the Floquet Hamiltonian takes the form $\hat{H}_F = 4E_C(\hat{n}-n_g)^2 - E_J\cos(\hat{\varphi} + \varphi_d\sin\hat{\vartheta})  + \hbar\omega_d\hat{m}$. In order to implement this Hamiltonian numerically, we expand the cosine term as $\cos(\hat{\varphi} + \varphi_d\sin\hat{\vartheta}) = \cos(\hat{\varphi})\cos(\varphi_d\sin\hat{\vartheta}) - \sin(\hat{\varphi})\sin(\varphi_d\sin\hat{\vartheta})$, and then use the Jacobi-Anger expansion to write
\begin{align}
\cos(\varphi_d\sin\hat{\vartheta}) &= J_0(\varphi_d) + 2\sum_{n=1}^\infty J_{2n}(\varphi_d)\cos(2n\hat{\vartheta}),\\
 \sin(\varphi_d\sin\hat{\vartheta}) &= 2\sum_{n=1}^\infty J_{2n - 1}(\varphi_d)\sin((2n - 1)\hat{\vartheta}),
\end{align}
where $J_\nu$ are Bessel functions of the first kind. We likewise perform such expansions for the tunneling operators to construct the perturbation Hamiltonian in Eq.~\eqref{eq:Gamma_abn_eval}.

\subsection{State labeling}
A technical difficulty that one encounters in Floquet numerical diagonalizations is the ambiguity of assigning state labels to highly hybridized modes. We can approach this using a dressed-basis technique where we track the driven eigenstates as the drive amplitude $\Omega$ is increased, similar to branch analysis and other related frameworks introduced in Refs.~\cite{shillito2022dynamics, xiao2023diagrammatic, dumas2024measurement,  goto2025labeling}. We start at $\Omega = 0$, where the Floquet eigenstates can be labeled unambiguously as product states $\ket{\alpha, m} = \ket{\alpha}\ket{m}$, and  then follow the eigenstates continuously to identify labels for the driven eigenstates $\ket{\tilde{\Phi}_{\alpha, m}} \equiv \ket{\widetilde{\alpha, m}}$. We have developed a highly numerically efficient implementation of this general state labeling algorithm in \texttt{jaxquantum} \cite{jha2024jaxquantum}, which is built using the \textsc{JAX} library in \textsc{Python} \cite{jax2018github}. Specifically, in our algorithm, we calculate overlaps $|\ip{\lambda(\Omega)}{\alpha, m}|$ as a function of $\Omega$, where $\ket{\lambda(\Omega)}$ are the eigenstates returned from numerical diagonalization at each drive amplitude. We then identify the state indices for $\ket{\tilde{\Phi}_{\alpha, m}}$ by following the smooth curve $|\ip{\tilde{\Phi}_{\alpha, m}(\Omega)}{\alpha, m}|$ at each value of $\Omega$, where $|\ip{\tilde{\Phi}_{\alpha, m}(\Omega = 0)}{\alpha, m}| = 1$ by construction. If the driven eigenstates deviate significantly from the undriven ($\Omega = 0$) basis, the corresponding overlaps may become too small to follow accurately. To mitigate this, we implement a periodic restart in the labeling procedure by updating the reference basis, i.e., calculating subsequent overlaps $|\ip{\lambda(\Omega > \Omega^\ast)}{\tilde{\Phi}_{\alpha, m}(\Omega^\ast)}|$ against a previously labeled driven state at drive amplitude $\Omega^\ast$ whenever the overlaps drop below a specified threshold.

\begin{figure}[t]
    \centering    
    \includegraphics[width=\linewidth]{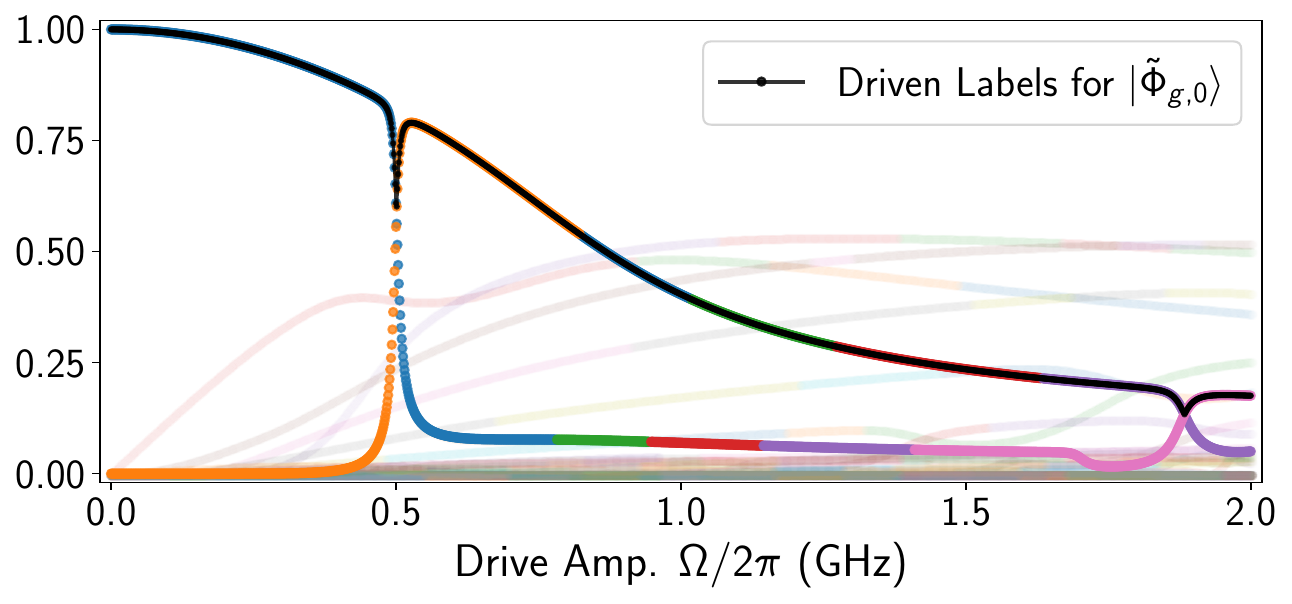}
    \caption{\textbf{Identifying labels for the Floquet eigenstates.} State overlaps $|\ip{\lambda(\Omega)}{g, 0}|$ as a function of the drive amplitude $\Omega$. Here $\ket{\lambda(\Omega)}$ are eigenstates obtained from numerical diagonalization of Eq.~\eqref{eq:HF_dressed_mode} for a driven transmon with parameters $E_J/h = 30$ GHz, $E_C/h = 0.15$ GHz, $n_g = 0$, and $\omega_d/2\pi = 5.059$ GHz, while $\ket{g, 0}$ is the undriven qubit ground state with Floquet photon-number index $m = 0$. The colors in this plot indicate the indices of the eigenstates returned from numerical diagonalization and the black solid curve represents the identified labels for the driven eigenstates $\ket{\tilde{\Phi}_{g, 0}} \equiv \ket{\widetilde{g, 0}}$. The two features at $\Omega/2\pi = 0.5$ GHz and $\Omega/2\pi = 1.9$ GHz reflect multiphoton resonances in the Floquet spectrum, which we intentionally cross diabatically in our labeling procedure.}
    \label{fig:5_state_labeling}
\end{figure}

As with all state assignment methods, our labeling procedure gets complicated by the presence of multiphoton resonances (i.e., avoided crossings) in the Floquet spectrum. At such resonances, assigning quantum numbers in the undressed basis in inherently ambiguous due to the hybridization between the involved states. We intentionally cross these resonances diabatically in our labeling method, as suggested in Ref.~\cite{weinberg2017adiabatic}. Accordingly, the curve $|\ip{\tilde{\Phi}_{\alpha, m}(\Omega)}{\alpha, m}|$ is constructed to remain smooth, with discontinuities only at multiphoton resonances. An illustrative example of this is shown in Fig.~\ref{fig:5_state_labeling}. Lastly, we note that it would also be possible to follow an adiabatic labeling using our state assignment algorithm if needed.

\subsection{Numerical simulation details\label{sec:sim_details}}

In this appendix, we provide additional details about the numerical simulations carried out in this work. The core aspect of the numerics involves diagonalizing and labeling the eigenstates $\ket{\tilde{\Phi}_{\alpha, m}}$ of the Floquet Hamiltonian $\hat{H}^F$ for the driven qubit at different values of the drive amplitude and frequency $\omega_d$ [cf. Equations \eqref{eq:H_transmon_disp_frame} and \eqref{eq:H_transmon_Rabi_frame}], and then computing the relevant matrix elements and frequencies to use in Equations~(\ref{eq:Gamma_abn_eval}--\ref{eq:Gamma_ab_eval}). 

For the transmon simulations in Figs.~\ref{fig:2_transmon} and \ref{fig:3_readout}, we set up the Hamiltonian of the undriven qubit in the charge basis using a cutoff of 100 charge levels, and then diagonalize this Hamiltonian to obtain the undriven energies and eigenstates. We then rotate the transmon operators $\hat{n}$ and $\cos(\hat{\varphi})$ into the eigenbasis and truncate further to reduce the numerical Hilbert space size, keeping only the $d$ lowest energy states. Using the transformed operators, we then set up the coupled-mode Floquet Hamiltonian as defined in Eq.~\eqref{eq:floquet_effective_HF_qubit} with a cutoff of $2M_{\rm max} + 1$ levels on the drive mode. The drive truncation $M_{\rm max}$ needs to be chosen such that the results of diagonalization converge, which in practice typically corresponds to $M_{\rm max} \gtrsim n$ with $n$ representing the largest $n$-photon process being considered. There are, however, numerical tricks one can utilize to reduce the drive Hilbert space required. For example, rather than computing matrix elements $\mel{\tilde{\Phi}_{\beta, -n}}{\hat{O}}{\tilde{\Phi}_{\alpha, 0}}$ as prescribed by Eq.~\eqref{eq:Gamma_abn_eval} directly, we can instead compute $\mel{\tilde{\Phi}_{\beta, m-n}}{\hat{O}}{\tilde{\Phi}_{\alpha, m}}$ with an initial Floquet photon-number index $m > 0$, since the Floquet replicas are identical. By applying this technique, we can nearly double the usable part of the Floquet Hilbert space [cf. Eq.~\eqref{eq:HF_matrix}]. In Figs.~\ref{fig:2_transmon} and \ref{fig:3_readout}, we used $M_{\rm max} = 15$ and a transmon truncation $d$ of 20 and 30 respectively, giving a total Hilbert space dimension $d \times (2M_{\rm max} + 1)$ between 600 and 900. 

In Fig.~\ref{fig:4_floquet0pi}, we similarly set up the SQUID Hamiltonian in the charge basis. However, here we did not transform into the energy eigenbasis and further truncate, since the emergent $\cos(2\hat{\varphi})$ potential can only be seen in the charge basis. We therefore kept all $d = 100$ charge states for the qubit. 
Furthermore, given the strong drives required, we found it necessary to use a drive truncation of $M_{\rm max} = 35$ to achieve convergence of the results. This resulted in a total Hilbert space dimension of 7100. Although the numerical Hilbert spaces used in this work are relatively large, we achieved fast simulation speeds by leveraging \texttt{jaxquantum} \cite{jha2024jaxquantum} to batch over drive parameters and perform parallelized parameter sweeps, as well as automatic state labeling.

Finally, an important aspect of our numerical approach is the use of distinct even- and odd-charge bases for diagonalization, allowing us to construct separate Hamiltonians for the two charge-parity sectors. We do this by defining the charge operator in the charge basis either as $\hat{n}_{\rm even}$ with integer values $k$ along the diagonal or as $\hat{n}_{\rm odd}$ with half-integer values $k + 1/2$. For each value of drive amplitude and frequency, we actually diagonalize these two charge-parity Hamiltonians separately, and compute matrix elements $\mel{\tilde{\Phi}_{\beta, m-n}^{\rm odd}}{\hat{O}}{\tilde{\Phi}_{\alpha, m}^{\rm even}}$, where $\hat{O}$ is a transition operator proportional to either $\cos(\hat{\varphi}/2)$ or $\sin(\hat{\varphi}/2)$. These operators, in turn, are numerically constructed in the undriven even $\to$ odd basis and explicitly defined to swap between the two parity sectors.

\section{DRIVE-ENHANCED QP TUNNELING \label{app:C_Conventional_QPT}}
In the main text, we investigated multiphoton-assisted QP generation due to strong microwave driving. In doing so, we ignored the effect of the drive on any pre-existing QPs in the device by assuming $x_{\rm qp} = 0$, i.e., a vanishing density of such QPs. Nonetheless, the theoretical framework we developed in Sec.~\ref{sec:2_Theoretical_Model} could equally well be applied to treat drive-induced modifications to the conventional QP tunneling process. In this appendix, we will comment on how such a treatment may be carried out.

The tunneling Hamiltonian perturbation from Eq.~\eqref{eq:H_T_QP_and_PAT} consists of two terms: one proportional to $\hat{\gamma}_{r s}^\dagger\hat{\gamma}_{l s} + {\rm h.c.}$, corresponding to conventional QP tunneling, and another proportional to $\hat{\gamma}_{r s}^\dagger\hat{\gamma}_{l \bar{s}}^\dagger + {\rm h.c.}$, corresponding to QP generation. In the main text, we focused solely on the second term by setting the QP density $x_{\rm qp} = 0$ and applying Fermi's golden rule to the generation process. In contrast, here we assume a finite $x_{\rm qp}$ and neglect the generation term, considering only conventional QP tunneling.

Under these assumptions, the tunneling Hamiltonian in Eq.~\eqref{eq:H_T_QP_and_PAT} (for a charge-driven transmon) simplifies to:
\begin{align}
\begin{split}
    \!\!&\hat{H}_T(t) \,\rightarrow\,\, \tau \sum_{l, r, s}\bigg[\left(u_r u_l-v_r v_l\right) \cos\bigg(\frac{\hat{\varphi} + \varphi_d(t)}{2}\bigg) \\ &\quad\,\,\,+i\left(u_r u_l+v_r v_l\right) \sin \bigg(\frac{\hat{\varphi} + \varphi_d(t)}{2}\bigg)\bigg]\hat{\gamma}_{r s}^{\dagger} \hat{\gamma}_{ls} + {\rm h.c.}\!\!\!\!
\end{split}
\label{eq:app_H_QP_only}
\end{align}
Here, $\varphi_d(t) = \varphi_d\sin(\omega_d t)$ is the oscillating phase induced across the JJ, as before, and the structure of Eq.~\eqref{eq:app_H_QP_only} mirrors that of Eq.~\eqref{eq:H_PAT_only} from the main text. We also remark that Eq.~\eqref{eq:app_H_QP_only} matches the perturbation Hamiltonian used in previous studies of QP tunneling in the absence of a drive~\cite{catelani2011relaxation, catelani2012decoherence, connolly2024coexistence}, apart from the replacement $\cos(\hat{\varphi}/2) \to \cos([\hat{\varphi} + \varphi_d\sin(\omega_d t)]/2)$ and similarly for the sine term. Thus, we can adapt these prior derivations to calculate the QP tunneling rates in the driven case:
\begin{align}
    % \begin{split}
         &\!\!\!\Gamma_{\alpha\beta}^{\rm qp} = \nonumber \\ &\!\!\sum_n\frac{16E_J}{h}\bigg[\big|\bra{\tilde{\Phi}_{\beta, n}}\cos\bigg(\frac{\hat{\varphi} + \varphi_d\sin\hat{\vartheta}}{2}\bigg)\ket{\tilde{\Phi}_{\alpha, 0}} \big|^2 \, \mathcal{S}_{\rm qp}^{-}\big(\omega_{\beta\alpha}^{(n)}\big) \nonumber\\
         &+\big|\bra{\tilde{\Phi}_{\beta, n}}\sin\bigg(\frac{\hat{\varphi} + \varphi_d\sin\hat{\vartheta}}{2}\bigg)\ket{\tilde{\Phi}_{\alpha, 0}} \big|^2 \mathcal{S}_{\rm qp}^{+}\big(\omega_{\beta\alpha}^{(n)}\big)\bigg].
    % \end{split}
    \label{eq:rate_QPT}
\end{align}
Here, $\omega_{\beta\alpha}^{(n)} =  n\omega_d + (\tilde{E}_{\beta, 0} - \tilde{E}_{\alpha, 0})/\hbar$ is the relevant transition frequency. Given the explicit summation over $n$, the transition rate above has already been projected back to the qubit Hilbert space from the Floquet extended space. Expressions for the QP tunneling structure factors  can be found in Refs.~\cite{catelani2011relaxation, catelani2012decoherence, connolly2024coexistence}; in particular, following Ref.~\cite{connolly2024coexistence}, we may decompose $\mathcal{S}_{\rm qp}^{\pm}$ into contributions from QPs tunneling across the JJ from the left lead to the right lead and vice versa: $\mathcal{S}_{\rm qp}^{\pm}(\omega) = \mathcal{S}_{\rm qp}^{L\to R, \,\pm}(\omega) + \mathcal{S}_{\rm qp}^{R\to L, \,\pm}(\omega)$. The structure factors depend on the temperature $T$, the QP density $x_{\rm qp}$, and, in the case of asymmetric gaps $\Delta_L$ and $\Delta_R$ on either side of the JJ, the total gap $2\bar{\Delta} \equiv \Delta_L + \Delta_R$ and the gap asymmetry $\delta\Delta \equiv \Delta_L - \Delta_R$. In the undriven case ($\varphi_d \to 0$), the only nonzero term in Eq.~\eqref{eq:rate_QPT} will be that with $n = 0$, and the expression for $\Gamma_{\alpha\beta}^{\rm qp}$ reduces to that found in prior studies. Turning on the drive then introduces additional contributions to QP-induced decoherence, the relevance of which will depend on the interplay between $\omega_d$, $\delta\Delta$, $\bar{\Delta}$, and the dressed qubit frequency.

% \clearpage
% Bibliography
\bibliography{bibliography}

\end{document}

\title{Supplementary note for \\ Theory of quasiparticle generation by microwave drives in superconducting qubits}
\author{Shoumik Chowdhury}
% \email{shoumikc@mit.edu}
\affiliation{Research Laboratory of Electronics, Massachusetts Institute of Technology, Cambridge, Massachusetts 02139, USA}
\affiliation{Department of Electrical Engineering and Computer Science, Massachusetts Institute of Technology, Cambridge, Massachusetts 02139, USA}
\author{Max Hays}
\affiliation{Research Laboratory of Electronics, Massachusetts Institute of Technology, Cambridge, Massachusetts 02139, USA}
\author{Shantanu R. Jha}
\affiliation{Research Laboratory of Electronics, Massachusetts Institute of Technology, Cambridge, Massachusetts 02139, USA}
\affiliation{Department of Electrical Engineering and Computer Science, Massachusetts Institute of Technology, Cambridge, Massachusetts 02139, USA}
\author{Kyle Serniak}
\affiliation{Research Laboratory of Electronics, Massachusetts Institute of Technology, Cambridge, Massachusetts 02139, USA}
\affiliation{MIT Lincoln Laboratory, Lexington, Massachusetts 02421, USA}
\author{\\ Terry P. Orlando}
\affiliation{Research Laboratory of Electronics, Massachusetts Institute of Technology, Cambridge, Massachusetts 02139, USA}
\affiliation{Department of Electrical Engineering and Computer Science, Massachusetts Institute of Technology, Cambridge, Massachusetts 02139, USA}
\author{Jeffrey A. Grover}
\affiliation{Research Laboratory of Electronics, Massachusetts Institute of Technology, Cambridge, Massachusetts 02139, USA}
\author{William D. Oliver}
\affiliation{Research Laboratory of Electronics, Massachusetts Institute of Technology, Cambridge, Massachusetts 02139, USA}
\affiliation{Department of Electrical Engineering and Computer Science, Massachusetts Institute of Technology, Cambridge, Massachusetts 02139, USA}
\affiliation{Department of Physics, Massachusetts Institute of Technology, Cambridge, Massachusetts 02139, USA}

% =====================================================
%%%%%%%%%%%%%%%%%%%%%%%%%%%%%%%%%%%%%%%
%%%%%%%%%%%%%% ABSTRACT %%%%%%%%%%%%%%
%%%%%%%%%%%%%%%%%%%%%%%%%%%%%%%%%%%%%%%
\date{\today}

\maketitle

\onecolumngrid

% =====================================================
%%%%%%%%%%%%%%%%%%%%%%%%%%%%%%%%%%%%%%%
%%%%%%%%%%%%%% MAIN TEXT %%%%%%%%%%%%%%
%%%%%%%%%%%%%%%%%%%%%%%%%%%%%%%%%%%%%%%

\section{Deriving the Josephson Junction Hamiltonian from BCS Theory\label{supp_sec1}}
A Josephson tunnel junction (JJ) consists of two superconductors separated by a thin insulating barrier through which Cooper pairs can quantum mechanically tunnel ~\cite{makhlin2001quantum, devoret2004superconducting, clarke2008superconducting, devoret2013superconducting, girvin2014circuit}. In this section, we provide a microscopic derivation of the Josephson Hamiltonian directly from BCS theory. Given its relevance to our work, this calculation is included here for completeness, since the microscopic derivation of the Josephson effect via a tunneling Hamiltonian is not typically presented in modern references on superconducting qubits and circuit quantum electrodynamics \cite{krantz2019quantum, kjaergaard2020superconducting, blais2020quantum}. In the steps that follow, we adapt and expand upon Refs. \cite{grabert1992single, martinis2004superconducting, girvin2019modern, glazman2021bogoliubov}. To begin, we consider a JJ that is described by the following BCS Hamiltonian:
\begin{align}
\hat{H} &= \hat{H}_L(\phi_L) + \hat{H}_R(\phi_R) + \hat{H}_T \label{eq:supp_H_BCS_tilde}\\ &= \sum_{l, s} \xi_l \hat{c}_{ls}^\dagger\hat{c}_{ls} + \sum_{l} \Delta_L\Big[ e^{i\phi_L} c_{l\uparrow}^\dagger c_{-l\downarrow}^\dagger + {\rm h.c.}\Big] +\, \sum_{r, s} \xi_r \hat{c}_{rs}^\dagger\hat{c}_{rs} + \sum_{r} \Delta_R\Big[ e^{i\phi_R} c_{r\uparrow}^\dagger c_{-r\downarrow}^\dagger + {\rm h.c.}\Big] +\, \tau\sum_{l,r,s}\Big[\hat{c}_{rs}^\dagger\hat{c}_{ls} + {\rm h.c.}\Big] \nonumber
\end{align}
Here, $\hat{H}_L$ and $\hat{H}_R$ are the mean-field BCS Hamiltonians of the superconducting leads on the left and right of the JJ, while $\hat{H}_T$ describes the tunneling of single electrons from the left to the right, and vice versa, with small tunneling amplitude $\tau$ \cite{glazman2021bogoliubov}. The BCS Hamiltonians $\hat{H}_L$ and $\hat{H}_R$ each consist of a free-electron term with second-quantized electronic energies $\xi_l$ and $\xi_r$ respectively (each referenced to their respective chemical potentials), as well as a pairing term proportional to the superconducting gap ($\Delta_L$ and $\Delta_R$ for the two leads, respectively). Note that $l$ and $r$ index electron momentum states on the left and right sides of the JJ, respectively, and thus $\hat{c}_{ls}$ is the fermionic annihilation operator for an electron\footnote{What we refer to as electrons here are technically themselves not ``bare'' electrons, but rather Landau quasiparticles consisting of a bare electron dressed by a positive background charge in a metal \cite{girvin2019modern}. However, as is typical in condensed matter systems, we continue to refer to them as electrons. We emphasize that these electronic Landau quasiparticles are distinct from the BCS quasiparticle excitations (QPs) that describe broken Cooper pairs in a superconductor. It is the generation of these BCS QPs that is main subject of this work.} with state $l$ and spin $s \in \{\uparrow, \downarrow\}$ in the left superconductor. Finally, the phases $\phi_L$ and $\phi_R$ are the superconducting phase order parameters for each superconductor. 

In order to proceed, it is helpful to perform a gauge transformation on Eq.~\eqref{eq:supp_H_BCS_tilde} so as to remove the superconducting phases $\phi_L$ and $\phi_R$ in the Hamiltonians $\hat{H}_L$ and $\hat{H}_R$, and instead transfer these phases into the tunneling Hamiltonian. We specifically use the unitary
\begin{equation}
\hat{U} = \hat{U}_L\hat{U}_R, \quad\text{with}\,\,\hat{U}_\iota = \exp\bigg(\frac{-i\phi_\iota}{2}\hat{N}_\iota\bigg)\,\,\text{for}\,\,\iota\in\{L,R\}.
\label{eq:supp_ULUR}
\end{equation}
Note that $\hat{N}_L = \sum_{l\sigma} \hat{c}_{l\sigma}^\dagger\hat{c}_{l\sigma}$ and $\hat{N}_R = \sum_{r\sigma} \hat{c}_{r\sigma}^\dagger\hat{c}_{r\sigma}$ are the number operators for electrons on the left and right sides of the JJ. The above unitary transforms Eq.~\eqref{eq:supp_H_BCS_tilde} via $\hat{H} \mapsto \hat{U}\hat{H}\hat{U}^\dagger$ and specifically has the following effect on the electron number operators:
\begin{equation}
\hat{c}_{l\sigma} \mapsto \hat{U}_L \hat{c}_{l\sigma} \hat{U}_L^\dagger = \hat{c}_{l\sigma} e^{i\phi_L/2}\quad\text{and} \quad \hat{c}_{l\sigma}^\dagger \mapsto \hat{c}_{l\sigma}^\dagger e^{-i\phi_L/2},
\end{equation}
and likewise for the electron operators for right side of the JJ. After performing the gauge transformation, we have
\begin{align}
\hat{H} &= \hat{H}_L(0) + \hat{H}_R(0) + \hat{H}_T(\varphi) \label{eq:supp_H_BCS}\\ &= \sum_{l, s} \xi_l \hat{c}_{ls}^\dagger\hat{c}_{ls} + \sum_{l} \Delta_L\Big[c_{l\uparrow}^\dagger c_{-l\downarrow}^\dagger + {\rm h.c.}\Big] +\, \sum_{r, s} \xi_r \hat{c}_{rs}^\dagger\hat{c}_{rs} + \sum_{r} \Delta_R\Big[c_{r\uparrow}^\dagger c_{-r\downarrow}^\dagger + {\rm h.c.}\Big] + \tau\sum_{l,r,s}\Big[e^{i\varphi/2}\hat{c}_{rs}^\dagger\hat{c}_{ls} + {\rm h.c.}\Big], \nonumber
\end{align}
where $\hat{H}_L(0)$ and $\hat{H}_R(0)$ are the BCS Hamiltonians without any phases, and we have now defined the \textit{superconducting phase difference} across the JJ as $\varphi = \phi_L - \phi_R$. 

At this stage, $\hat{H}_L(0)$ and $\hat{H}_R(0)$ can be diagonalized exactly via a Bogoliubov transformation $\hat{c}_{ks} = u_k\hat{\gamma}_{ks} + \sigma(s) v_k\hat{\gamma}_{k\bar{s}}^\dagger$ where $\sigma(s) = \pm 1$ for $s = \,\uparrow, \downarrow$ and $\bar{s}$ denotes the opposite spin to $s$. Here, $\hat{\gamma}_{ls}$ ($\hat{\gamma}_{rs}$) is the fermionic annihilation operator for a \textit{BCS quasiparticle excitation} in the left (right) lead of the JJ, with the conventional BCS coherence factors $u_k, v_k$ defined via $|v_k|^2 = 1 - |u_k|^2 = (1 - \xi_k/\varepsilon_k)/2$ \cite{orlando1991foundations, tinkham2004introduction, catelani2011relaxation}. We note that we can work in a gauge where $u_k$ and $v_k$ are both real-valued. Using this Bogoliubov transformation, we can diagonalize the first two terms of Eq.~\eqref{eq:supp_H_BCS} above to get
\begin{equation}
    \hat{H}_{\rm QP} = \hat{H}_L(0) + \hat{H}_R(0) = E_{\rm GS} + \sum_{l, s} \varepsilon_l \hat{\gamma}_{ls}^\dagger \hat{\gamma}_{ls} + \sum_{r, s} \varepsilon_r \hat{\gamma}_{rs}^\dagger \hat{\gamma}_{rs}, \label{eq:supp_HQP}
\end{equation}
with energies $\varepsilon_l = \sqrt{\xi_l^2 + \Delta_L^2}$ and $\varepsilon_r = \sqrt{\xi_r^2 + \Delta_R^2}$ for the quasiparticle excitations (QPs) on the left and right sides of the JJ, respectively, and where $E_{\rm GS}$ is an overall energy for the joint BCS ground state of the two superconductors. We note that one important difference between Equations \eqref{eq:supp_H_BCS_tilde} and \eqref{eq:supp_H_BCS} is the treatment of the ground state. In Eq.~\eqref{eq:supp_H_BCS_tilde}, there is no unique ground state of $\hat{H}_L(\phi_L) + \hat{H}_R(\phi_R)$ [the BCS Hamiltonians], but rather an entire \textit{ground state manifold} spanned by a family of degenerate states that can be parameterized by the phase difference $\varphi = \phi_L -\phi_R$. Alternatively, one can parameterize the states in this manifold via an integer $n$ which counts the number of Cooper pairs that have tunneled across the JJ. In the picture of Eq.~\eqref{eq:supp_H_BCS_tilde}, $\hat{H}_T$ then breaks the degeneracy within the ground state manifold. By contrast, the joint ground state of $\hat{H}_L(0) + \hat{H}_R(0)$ in Eq.~\eqref{eq:supp_H_BCS} is independent of the superconducting phases and thus unique, with the $\varphi$-dependence instead shifted to $\hat{H}_T$. It can be expressed as the product of the BCS ground states for each individual superconductor: $\ket{{\rm GS}} = \ket{{\rm BCS}_L}\otimes\ket{{\rm BCS}_R}$, where both constituent ground states have their phases set to zero in this gauge \cite{glazman2021bogoliubov}. Therefore, $\ket{{\rm BCS}_L} = \prod_l (u_l + v_l c_{l\uparrow}^\dagger c_{-l\downarrow}^\dagger)\ket{{\rm vac}}$ and likewise for $\ket{{\rm BCS}_R}$, with $\ket{{\rm vac}}$ being the vacuum state for electrons. 

We are now ready to perform perturbation theory on Eq.~\eqref{eq:supp_H_BCS} in order to derive the Josephson potential. Specifically, we treat $\hat{H}_T(\varphi)$ as a perturbation with small parameter $\tau$, and write $\hat{H} = \hat{H}_{\rm QP} + \hat{H}_T(\varphi)$. It is convenient to express $\hat{H}_T$ in the form
\begin{equation}
    \hat{H}_T(\varphi) = \hat{T}_+ e^{i\varphi/2} + \hat{T}_- e^{-i\varphi/2}, \quad\text{with}\,\,\, \hat{T}_+ = \tau\sum_{l, r, s}\hat{c}_{rs}^\dagger\hat{c}_{ls}\,\,\,\text{and}\,\,\, \hat{T}_- = (\hat{T}_+)^\dagger,
    \label{eq:supp_HT_splitup}
\end{equation}
which makes readily apparent the fact that $\mel{{\rm GS}}{\hat{H}_T}{{\rm GS}} = 0$. Thus, we expect that $\hat{H}_T$ will not produce any correction to the ground state energy at first order. Intuitively, this is not surprising since $\hat{H}_T$ transfers single-electrons across the JJ. To capture the effect of Cooper-pair tunneling---which requires two successive single-electron tunneling events---we must therefore evaluate the second-order energy shift:
\begin{equation}
    \delta E_{\rm GS}^{(2)} = -\sum_{k \neq {\rm GS}} \frac{|\mel{k}{\hat{H}_T(\varphi)}{\rm GS}|^2}{E_k - E_{\rm GS}} = -\mel{{\rm GS}}{\hat
{H}_T(\varphi)\hat{\mathcal{A}}\hat{H}_T(\varphi)}{{\rm GS}},\quad\text{with}\,\,\, \hat{\mathcal{A}} = \sum_{k \neq {\rm GS}}\frac{\op{k}{k}}{E_k - E_{\rm GS}}.
\end{equation}
Here, the intermediate states $\ket{k}$ are excited eigenstates of $\hat{H}_{\rm QP}$ with associated energies $E_k$. Plugging in Eq.~\eqref{eq:supp_HT_splitup}, we get
\begin{equation}
    \delta E_{\rm GS}^{(2)} = -\Big[e^{i\varphi}\mel{{\rm GS}}{\hat{T}_+ \hat{\mathcal{A}} \hat{T}_+}{{\rm GS}} + e^{-i\varphi}\mel{{\rm GS}}{\hat{T}_- \hat{\mathcal{A}} \hat{T}_-}{{\rm GS}} + \mel{{\rm GS}}{\hat{T}_+ \hat{\mathcal{A}} \hat{T}_-}{{\rm GS}} + \mel{{\rm GS}}{\hat{T}_- \hat{\mathcal{A}} \hat{T}_+}{{\rm GS}}\Big].
\end{equation}
While all four terms have nonzero matrix element, the latter two are independent of $\varphi$ and therefore will only contribute an overall constant energy shift, which we can absorb into the unperturbed $E_{\rm GS}$. Meanwhile, given that $\hat{H}_T$ and $\ket{{\rm GS}}$ are real and $\hat{\mathcal{A}}$ is hermitian, we have $A \equiv \mel{{\rm GS}}{\hat{T}_+ \hat{\mathcal{A}} \hat{T}_+}{{\rm GS}} = A^\ast = \mel{{\rm GS}}{\hat{T}_- \hat{\mathcal{A}} \hat{T}_-}{{\rm GS}}$. Therefore, up to constants, we find that
\begin{equation}
    \delta E_{\rm GS}^{(2)} = -A\big[e^{i\varphi} + e^{-i\varphi}\big] = -E_J\cos(\varphi),
\end{equation}
where the exact expression for $E_J = 2A$ reduces to the standard Ambegaokar-Baratoff relation \cite{ambegaokar1963tunneling} upon evaluating the matrix element $A$ (which is done explicitly in Refs. \cite{martinis2004superconducting, girvin2019modern}). We have thus derived the Josephson potential as the second-order perturbative contribution of the tunneling Hamiltonian to the BCS ground state. Note that while the calculation was carried out here using the language of energy shifts and perturbation theory, it can also equivalently be done at the Hamiltonian level, using a Schrieffer-Wolff transformation to define an effective Hamiltonian in a low-energy subspace (which, here, is the ground state manifold) \cite{liao2024circuit}. Finally, as discussed in Appendix A of the main article, we may promote $\varphi$ to a fully quantum degree of freedom $\hat{\varphi}$, in which case the effective JJ Hamiltonian in the ground state manifold is $\hat{U} = -E_J\cos(\hat{\varphi})$. 

\section{Microwave Drives and Charging Effects\label{supp_sec2}}

In this section, we will build on the derivation of the Josephson potential presented in Sec.~\ref{supp_sec1}, and further incorporate both the charging energy for finite-size superconducting islands, as well as the effect of microwave driving. Our starting Hamiltonian is a modified version of Eq.~\eqref{eq:supp_H_BCS_tilde}. Without loss of generality, we take the right lead of the JJ to be grounded, in which case
\begin{equation}
   \hat{H} = \hat{H}_L\Big(\phi_L + \int dt\, 2eV(t)/\hbar\Big) - eV(t)\hat{N}_L + E_C(\hat{N}_L - \mathcal{N}_g)^2 + \hat{H}_R(\phi_R) + \hat{H}_T.
   \label{eq:supp_H_BCS_charging_microwave}
\end{equation}
Here, $\hat{H}_R$ and $\hat{H}_T$ are the same as in Eq.~\eqref{eq:supp_H_BCS_tilde}, whereas the Hamiltonian for the left lead has been modified to incorporate an oscillating voltage $V(t)$ across the JJ via $\hat{H}_L(\phi_L) \to \hat{H}_L(\phi_L + \int dt\, 2eV(t)/\hbar) - eV(t)\hat{N}_L$ \cite{grabert1992single}. We furthermore include the electrostatic energy of the left island, which depends on the single-electron charging energy $E_C = e^2/2(C_J + C_g)$, and the electron dc offset charge $\mathcal{N}_g = C_gV_g/e$. Here, $C_J$ and $C_g$ are the intrinsic JJ capacitance and the capacitance between the island and a gate electrode with gate voltage $V_g$, respectively. 

The modification of $\hat{H}_L$ above is required to be consistent with the second Josephson relation that relates the phase difference between the leads to the time-dependent voltage $V(t)$ across the JJ. The specific source of this oscillating voltage is not important here --- it could come from a direct voltage bias, irradiation by microwaves, or the induced voltage due to a time-dependent gate offset $V_g(t)$ that results
in a voltage $V(t) = C_g V_g(t)/(C_J + C_g)$ across the JJ (in this last situation, any dc component of the gate voltage enters via the static charging term, while the ac component is what contributes to the instantaneous voltage $V(t)$ across the JJ). In all cases, this drive-induced voltage $V(t)$ enters into the Hamiltonian in two places [as given in Eq.~\eqref{eq:supp_H_BCS_charging_microwave}], and therefore couples to both the phase variable and to the electrons themselves via $\hat{N}_L$. 

We can now proceed as before by removing the phase dependence from the individual BCS Hamiltonians and shifting it to the tunneling Hamiltonian. We first apply a time-dependent unitary $\hat{W}(t) = \exp[i\int eV(t)\hat{N}_L/\hbar dt]$ which both removes the time-dependent phase of the left lead and cancels out the term $-eV(t)\hat{N}_L$. We then apply the unitary $\hat{U}$ from Eq.~\eqref{eq:supp_ULUR}. Both of these unitaries leave the charging energy term unaffected, and the resulting Hamiltonian is
\begin{equation}
   \hat{H} = \hat{H}_L(0) + \hat{H}_R(0) +  E_C(\hat{N}_L - \mathcal{N}_g)^2 +  \tau\sum_{l,r,s}\Big[e^{i(\varphi + \varphi_d(t))/2}\hat{c}_{rs}^\dagger\hat{c}_{ls} + {\rm h.c.}\Big],
\end{equation}
where we have explicitly written down the time-dependent tunneling Hamiltonian $\hat{H}_T$ in terms of the phase difference $\varphi$ and the additional oscillating phase $\varphi_d(t) = \int dt\, 2eV(t)/\hbar$ due to the drive.

At this stage we can again perform second-order perturbation theory, now treating both the charging term and $\hat{H}_T$ as perturbations, and then promote the phase difference to a quantum operator $\hat{\varphi}$. The derivation of the Josephson potential proceeds exactly as before, and now results in a \textit{time-dependent} Josephson potential $\hat{U} = -E_J\cos(\hat{\varphi} + \varphi_d(t))$. Meanwhile, the single-electron charging term can be replaced by an equivalent Cooper pair charging term $4E_C(\hat{n} - n_g)^2$, where $\hat{n}$ represents the number of Cooper pairs that have tunneled across the JJ and satisfies the commutation relation $[\hat{n}, e^{i\hat{\varphi}}] = e^{i\hat{\varphi}}$. Explicitly, we define $\hat{n} = (\hat{N}_L - \hat{N}_R)/2$ and note that $\hat{N}_{\rm tot} = \hat{N}_L + \hat{N}_R$ is a constant since $[\hat{H}, \hat{N}_{\rm tot}] = 0$. Thus, in making the replacement above, we absorb the constant mean electron number $N_{\rm tot}/2$ into $\mathcal{N}_g$ and then define the Cooper pair offset charge $n_g = \mathcal{N}_g/2 = C_gV_g/2e$ in the conventional way. 

Finally, in order to set ourselves up to perform the \textit{first-order} perturbation theory calculation for QP generation in the main article, we can write down the full system Hamiltonian as
\begin{equation}
    \hat{H} = 4E_C(\hat{n} - n_g)^2 - E_J\cos(\hat{\varphi} + \varphi_d(t)) + \hat{H}_{\rm QP} +  \bigg\{\tau\sum_{l,r,s}\Big[e^{i(\varphi + \varphi_d(t))/2}\hat{c}_{rs}^\dagger\hat{c}_{ls} + {\rm h.c.}\Big] + E_J\cos(\hat{\varphi} + \varphi_d(t))\bigg\}.
    \label{eq:supp_full_Hamiltonian_mainEq1}
\end{equation}
The first two terms here represent the Hamiltonian for the superconducting qubit arising from the JJ above, while $\hat{H}_{\rm QP} = \hat{H}_L(0) + \hat{H}_R(0)$ represents the QPs in the two superconducting leads. The tunneling Hamiltonian $\hat{H}_T(t)$, written in parentheses above, is time-dependent and furthermore now incorporates a correction $E_J\cos(\hat{\varphi} + \varphi_d(t))$ to avoid double counting the Josephson energy term \cite{catelani2011relaxation}. This is important since the Josephson potential in the qubit Hamiltonian itself arises from the tunneling term at second-order, as we have seen here in this note. Equation \eqref{eq:supp_full_Hamiltonian_mainEq1} is exactly the starting Hamiltonian of the main text, and correctly incorporates the time-dependence of the microwave drive on both the superconducting qubit degree of freedom and the QPs themselves.

% Bibliography
% Before manual correction in Python
% \bibliography{bibliography}

% After generating fixed bbl file
\providecommand{\noopsort}[1]{}\providecommand{\singleletter}[1]{#1}%
%

% \clearpage